\documentclass[aps,prl,reprint,amsmath,superscriptaddress]{revtex4-2}
\usepackage{amsmath}
\usepackage{txfonts}
\usepackage{hyperref}
\usepackage{graphicx}
\usepackage{float}
\usepackage{xcolor}
\usepackage[normalem]{ulem}
\usepackage{todonotes}
\usepackage[utf8]{inputenc}
\usepackage{bm}
\usepackage{etoolbox}
\usepackage[caption=false]{subfig}

\makeatletter
\def\maketitle{
\@author@finish
\title@column\titleblock@produce
\suppressfloats[t]}
\makeatother

\DeclareMathOperator{\erf}{erf}
\newcommand{\qed}{\hfill\blacksquare}
\newcommand{\AllenAffiliation}{Allen Institute, Seattle, WA}
\newcommand{\Dz}{D z}

\begin{document}

\title{
Hierarchy of chaotic dynamics in random modular networks
}
\author{\L{}ukasz Ku\'smierz}
\affiliation{\AllenAffiliation}
\author{Ulises Pereira-Obilinovic}
\affiliation{\AllenAffiliation}
\author{Zhixin Lu}
\affiliation{\AllenAffiliation}
\author{Dana Mastrovito}
\affiliation{\AllenAffiliation}
\author{Stefan Mihalas}
\affiliation{\AllenAffiliation}

\begin{abstract}
We introduce a model of randomly connected neural populations 
and study its dynamics by means of 
the dynamical mean-field theory and simulations. 
Our analysis uncovers a rich phase diagram, featuring 
high- and low-dimensional chaotic phases, 
separated by a crossover region characterized by 
low values of the maximal Lyapunov exponent 
and participation ratio dimension, 
but with high values of the Lyapunov dimension that change significantly 
across the region. 
Counterintuitively, chaos can be attenuated by either 
adding noise to strongly modular connectivity  
or by introducing modularity into random connectivity. 
Extending the model to include a multilevel, 
hierarchical connectivity reveals that 
a loose balance between activities across levels 
drives the system towards the edge of chaos. 
\end{abstract}
\maketitle
    
\textit{Introduction.---}Natural and artificial systems at the edge of chaos exhibit unique computational properties~\cite{packard1988adaptation,langton1990computation,kauffman1991coevolution,mora2011biological}. 
For instance, placing artificial neural networks at the edge of chaos has been shown to optimize learning and information processing in various scenarios~\cite{bertschinger2004real,legenstein2007edge,sussillo2009generating,toyoizumi2011beyond,schoenholz2016deep,schuecker2018optimal}. 
Accordingly, it has been suggested that neural circuits in the brain operate at, 
or remain near to, a critical point such as the edge of chaos~\cite{herz1995earthquake,stassinopoulos1995democratic,chialvo1999learning,kinouchi2006optimal,beggs2008criticality,beggs2012being,munoz2018colloquium,fontenele2019criticality,castro2024in}, 
which may explain some aspects of the structure of the variability observed in neural recordings. 
However, keeping complex systems at a critical point is challenging, 
as it is usually sensitive to external or internal changes, such as  
noise and external inputs~\cite{molgedey1992suppressing,rajan2010stimulus,schuecker2018optimal,takasu2024suppression}, 
or low-rank structural connectivity perturbations \cite{rajan2006eigenvalue,cabana2013large,aljadeff2015transition,landau2018coherent,mastrogiuseppe2018linking,pereira2023forgetting,mastrovito2024transition,clark2024structure}. 
How can complex systems be robustly maintained at the edge of chaos? 
In the context of neuroscience, synaptic plasticity has been suggested 
as the underlying feedback mechanism driving neural circuits towards a critical point
\cite{de2006self,levina2007dynamical,shew2015adaptation,zeraati2021self}. 
It is unclear, however, whether synaptic plasticity mechanisms can drive networks specifically to the edge of chaos \cite{dahmen2019second} 
and how the underlying connectivity organization affects the type and 
robustness of the emergent critical behavior. 

Hierarchical organization is common in biological and artificial complex systems, including brains, ecosystems, and artificial neural networks. 
Mammalian brains are modular and hierarchical \cite{zhou2006hierarchical,he2009uncovering,meunier2009hierarchical,meunier2010modular,sporns2016modular,harris2019hierarchical}, 
and are composed of many distinct cell types that themselves are hierarchically organized \cite{tasic2018shared,zeng2022cell}. 
The hierarchical organization of the visual system \cite{felleman1991distributed} 
has inspired the design of computationally powerful artificial neural network 
architectures~\cite{lecun2015deep}, which are similarly hierarchically structured~\cite{Goodfellow-et-al-2016}. 
Ecological systems, including food webs, also exhibit hierarchical organization \cite{pascual2006ecological}. 
How does hierarchical organization affect the collective behavior of such complex systems? 
Simulation studies suggest that hierarchical modular connectivity broadens the parameter space supporting critical-like behavior~\cite{rubinov2011neurobiologically,wang2012hierarchical,moretti2013griffiths}, but the exact mechanism, its universality, and its relation to the edge of chaos remain unclear. 

Thus, we set out to understand how modular and hierarchical connectivity organization affects dynamics at different levels of the hierarchy. 
As we demonstrate below, the presence of modules gives rise to qualitatively distinct chaotic phases. 
High-dimensional chaotic activity (microscopic chaos) 
common in homogeneous networks 
\cite{sompolinsky1988chaos,clark2023dimension} 
is separated from the low-dimensional chaotic activity of strongly 
coherent modules (macroscopic chaos) 
by an interesting crossover region (multiscale chaos), 
wherein both forms of chaotic activity coexist 
and the dimension of activity can be interpreted 
as either high or low, depending on the measure used. 
Furthermore, random hierarchical interaction structures 
significantly enhance the robustness of the edge of chaos 
without the need for precise fine tuning. 
The underlying mechanism is remarkably simple: 
Different levels of the hierarchy effectively compete for activity 
and the hierarchical organization in the network's interactions 
coarse-grains chaotic fluctuations from lower levels to higher ones, 
preventing the amplification of chaos at higher levels in the hierarchy.

\textit{Model.---}We study a class of random network models commonly used in neuroscience~\cite{sompolinsky1988chaos}, artificial intelligence~\cite{sussillo2009generating, poole2016exponential}, and ecology~\cite{allesina2015predicting,roy2019numerical, garcia2024interactions}, incorporating modular and hierarchical organization into their interactions. We structure these interactions in levels, 
where higher levels are constructed by coarse-graining lower levels.
Despite the generality of the model, for concreteness we refer to nodes of the network 
as neurons and to interaction strengths between them as synaptic weights.  
We start by introducing and analyzing a two-level modular system, 
in which the interaction matrix is generated in 
blocks corresponding to distinct populations, 
i.e. clusters of neurons with similar connectivity patterns. 
In effect, populations are introduced through a random 
and relatively low-rank connectivity perturbation \cite{arbenz1988spectral,tao2013outliers} 
and, in the context of neuroscience, 
can be interpreted as emerging from shared spatial locations, 
morphologically or genetically defined cell types, 
or functions as in neuronal ensembles or engrams.  
Later we generalize our model and analysis by incorporating multiple levels. 

Since we focus our attention on the effects of connectivity patterns, 
we choose a simple rate model with dynamics shared across all populations:
\begin{equation}
    \boldsymbol{x}(t+1) 
    =
    \phi\left(
    \boldsymbol{J} \boldsymbol{x}(t)
    \right)
    \label{eq:model_dynamics}
\end{equation}
where $\boldsymbol{x}(t)$ is a vector of neural activity 
at time $t$, $\boldsymbol{J}$ is the connectivity matrix, 
and the activation function $\phi:\mathbb{R}\to\mathbb{R}$ is applied element-wise. 
We take $\boldsymbol{J}$ to be a block matrix of the form 
\begin{equation}
    \boldsymbol{J} 
    =
    \begin{bmatrix}
    \boldsymbol{J^{11}} & \boldsymbol{J^{12}} & \dots & \boldsymbol{J^{1P}} 
    \\
    \boldsymbol{J^{21}} & \boldsymbol{J^{22}} & \dots & \boldsymbol{J^{2P}} 
    \\
      \vdots          & \vdots          & \ddots & \vdots 
    \\
    \boldsymbol{J^{P1}} & \boldsymbol{J^{P2}} & \dots & \boldsymbol{J^{PP}} 
\end{bmatrix}
\end{equation} 
where each block $\boldsymbol{J^{\alpha\beta}}$ is a submatrix that 
contains all synaptic weights from neurons in 
population $\beta$ to all neurons in population $\alpha$. 
For simplicity of notation, here we limit 
our attention to $P$ populations of the same size $n$ 
\footnote{
This assumption could be easily relaxed. In general, each block could be rectangular, reflecting 
different sizes of populations. 
The results would not be affected as long as the number 
of neurons in each population is large enough to justify 
the validity of the mean-field approach utilized in our analysis.}.

\begin{figure}
    \centering
    \includegraphics[width=0.99\linewidth]{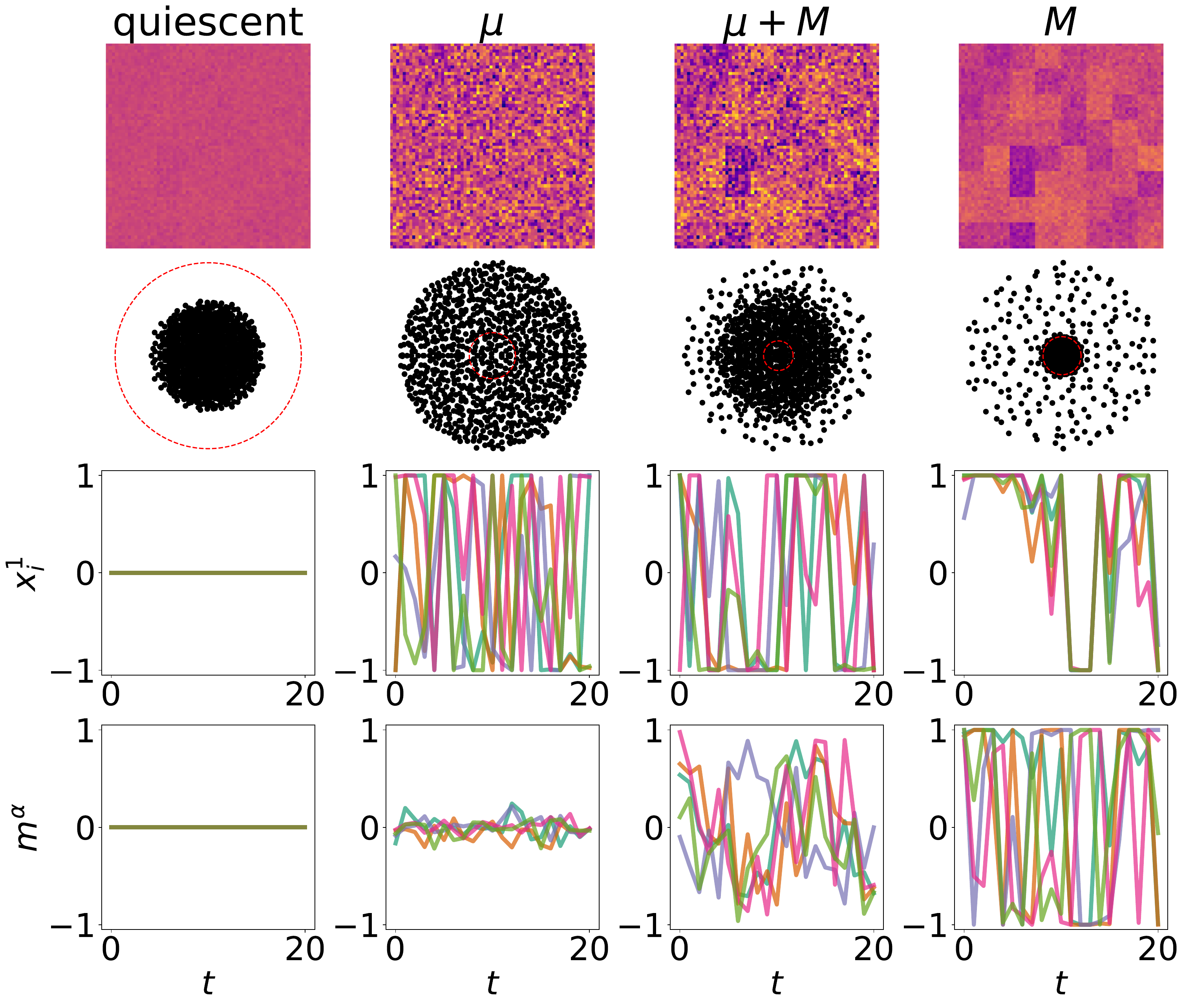}
    \caption{
    Visualizations of the weight matrix (\textit{top row}), 
    its eigenvalues (\textit{second row}), 
    and steady-state activities 
    of five sample neurons from a shared population (\textit{third row}) 
    and five sample populations (\textit{bottom row}) 
    in four phases predicted by the mean-field theory: 
    quiescent (${\sigma=\sigma_{\mu}=0.5}$), 
    $\mu$ (microscopic chaos, 
    ${\sigma=4, \sigma_{\mu}=0.5}$),
    $\mu+M$ (multiscale chaos, 
    ${\sigma=4, \sigma_{\mu}=6}$),
    and $M$ (macroscopic chaos,  
    ${\sigma=1, \sigma_{\mu}=5}$).
    Red dashed lines in the eigenvalue spectra represent 
    the unit circle in the complex plane.
    }
    \label{fig:eigvals_and_activities}
\end{figure}

To obtain a minimal and solvable model with characteristics 
of population-specific connectivity patterns, 
we define a hierarchical weight-generating process as follows. 
We endow each pair of populations $(\alpha, \beta)$  
with a pair of parameters 
$(\mu^{\alpha\beta}, \sigma^{\alpha\beta})$
which are used to generate random weights between neurons in these populations as
${J^{\alpha\beta}_{ij} \sim \mathcal{N}(\mu^{\alpha\beta}/n, \sigma^{\alpha\beta}/\sqrt{n})}$. 
These pairs of parameters are themselves generated randomly i.i.d. from 
a specific distribution. 
In this work we focus our attention on the case of 
${\sigma^{\alpha\beta}=\sigma/\sqrt{P}}$ 
fixed and shared across all pairs of populations 
and $\mu^{\alpha\beta}$ 
generated randomly as $
    \mu^{\alpha\beta} \sim 
    \mathcal{N}\left(0, \sigma_{\mu}/\sqrt{P}\right)
    $. 
The overall connectivity matrix can be concisely expressed as
\begin{equation}
    \boldsymbol{J} = 
    \sigma_{\mu} \boldsymbol{\Xi^{(P)}} \otimes \boldsymbol{O^{(n)}} + \sigma \boldsymbol{\Xi^{(N)}}
    \label{eq:J_clustered_flat}
\end{equation}
where $\otimes$ stands for the Kronecker product, 
$N=nP$ is the total number of neurons, 
${\boldsymbol{O^{(n)}} = v v^T}$ 
is a fixed $n\times n$ 
orthogonal projection matrix with 
${v^T = \frac{1}{\sqrt{n}}
    \begin{bmatrix}
1 & 1 & \dots & 1
\end{bmatrix}}$, 
and $\boldsymbol{\Xi^{(m)}}$ is a random 
$m \times m$ matrix with each 
entry generated i.i.d. from $\mathcal{N}(0, 1/\sqrt{m})$. 
Note that matrices $\boldsymbol{\Xi^{(P)}}$ and $\boldsymbol{\Xi^{(N)}}$ 
are mutually independent and correspond to different 
levels in the hierarchy of randomness: 
$\boldsymbol{\Xi^{(P)}}$ to mean efficacies between 
pairs of populations and $\boldsymbol{\Xi^{(N)}}$ to random fluctuations 
at the level of pairs of individual neurons.
The resulting connectivity matrix has a block 
structure with a double-disk distribution 
of eigenvalues \cite{tao2013outliers}, 
see Fig.~\ref{fig:eigvals_and_activities}.
\begin{figure*}
        \includegraphics[width=0.29\textwidth]{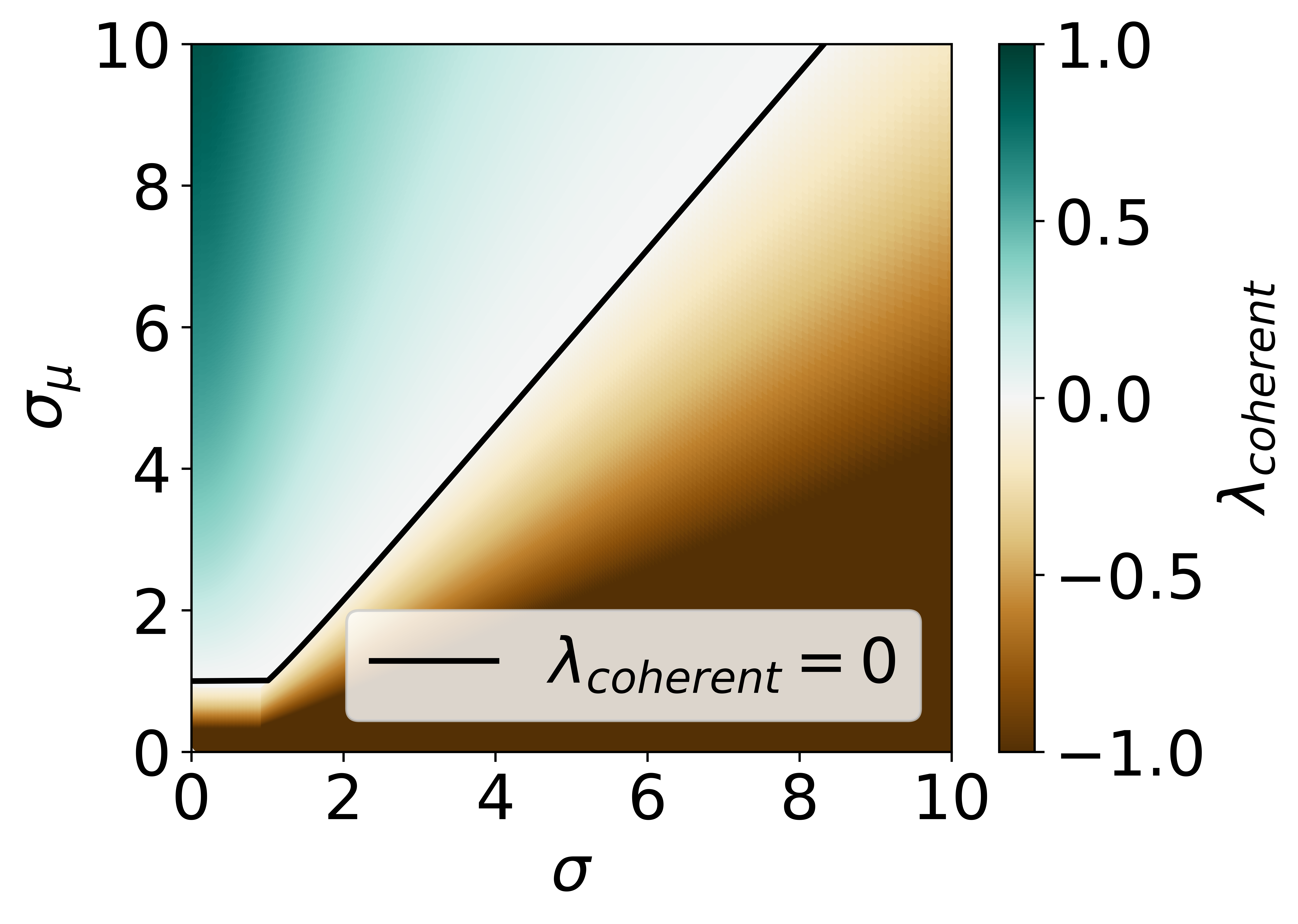}
        \includegraphics[width=0.29\textwidth]{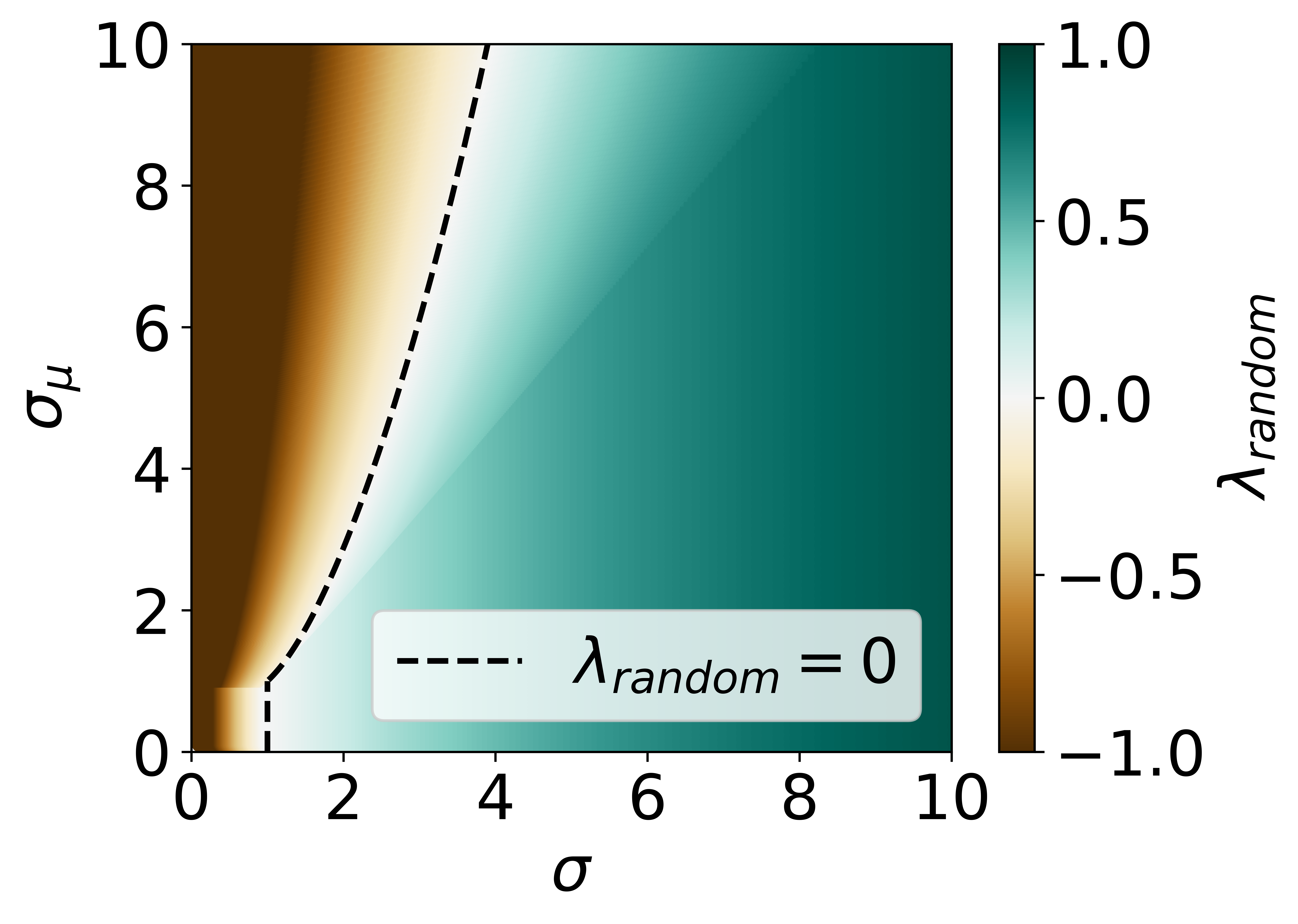}
        \includegraphics[width=0.29\textwidth]{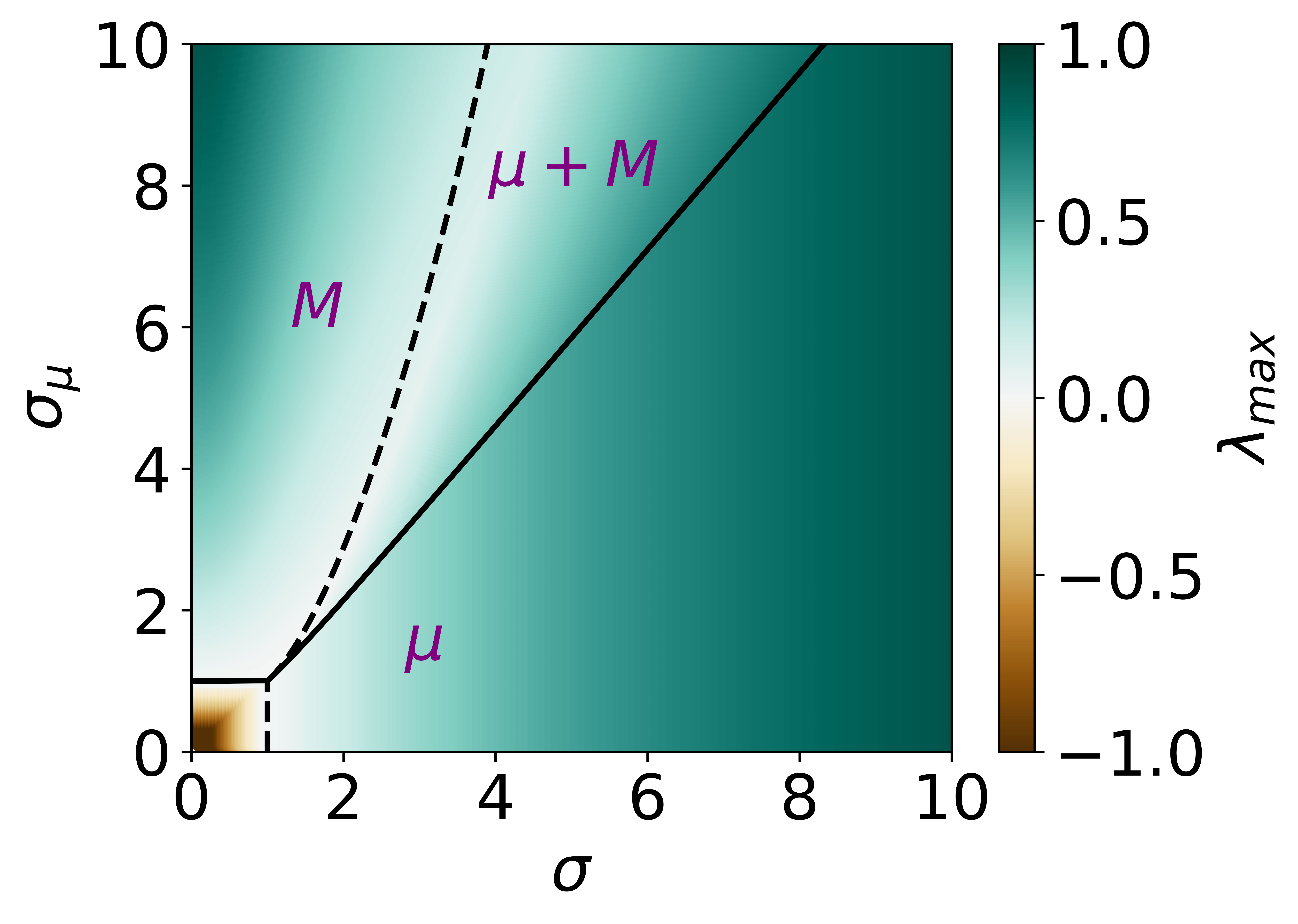}
    \caption{
    The mean-field phase diagram.  
    Lines represent transition lines: 
    $\sigma^*_{\mu}$ (solid) 
    and 
    $\sigma^{EoHC}_{\mu}$ (dashed). 
    Phases are denoted by symbols 
    $M$ (macroscopic chaos) and
    $\mu$ (microscopic chaos). 
    Heat maps represent the values of 
    $\lambda_{coherent}$ \textit{(left)},
    $\lambda_{random}$ \textit{(center)}, 
    and
    $\lambda_{max}$ \textit{(right)}. 
    }
    \label{fig:phase_diagram}
\end{figure*}

\textit{Mean-field analysis.---}In order to enable a straightforward mathematical 
analysis of the typical behavior of a network with multiple populations of neurons we consider a rather 
natural limit of a large number of large populations ${1 \ll P \ll }N$. 
The activity of neuron $i$ in population $\alpha$ 
evolves according to (\ref{eq:model_dynamics}), which 
we rewrite as
\begin{subequations}
    \label{eq:update_with_currents}
    \begin{align}    
    &x_i^{\alpha}(t+1)
    =
    \phi\left(
    h^{\alpha}_i(t)
    \right)
\\
    &h^{\alpha}_i(t)
    =
    \sum_{\beta = 1}^P \sum_{j=1}^{n}
    J_{ij}^{\alpha\beta}x_j^{\beta}(t)
    \end{align}
\end{subequations}
where the input currents (preactivations) can be expressed as
\begin{equation}
    h^{\alpha}_i(t)
    =
    \sum_{\beta = 1}^P \mu^{\alpha\beta} 
    m^{\beta}(t)
    +
     \sum_{\beta = 1}^P \sum_{j=1}^{n}
     \sigma \Xi^{\alpha \beta}_{ij}x_j^{\beta}(t)
     \label{eq:input_current_detailed}
\end{equation}
with i.i.d. $\Xi^{\alpha \beta}_{ij}\sim\mathcal{N}(0, 1/\sqrt{N})$ 
and where we define the mean activity of population $\alpha$ as
\begin{equation}
    m^{\alpha}(t) = \frac{1}{n} \sum_{i=1}^{n} x_i^{\alpha}(t)
    \label{eq:mean-population-activities}
\end{equation}
When considered as a function of quenched synaptic weight disorder, 
the input current of each neuron is a Gaussian random variable. 
To keep track of the correlation structure between 
different neurons we rewrite it as a sum of
population-specific and neuron-specific terms
\begin{equation}
    h^{\alpha}_i(t)
    =
    \sigma_{\mu} \sqrt{q_m(t)} \tilde{z}^{\alpha}
    +
    \sigma \sqrt{q(t)} z^{\alpha}_i
    \label{eq:current_with_fields}
\end{equation}
where $\tilde{z}^{\alpha}$, $z^{\alpha}_i$ are 
i.i.d. standard Gaussian random variables 
and 
\begin{subequations}
\label{eq:definition_order_parameters}
\begin{align}
    q_m(t) 
    &= 
    \frac{1}{P}\sum_{\alpha=1}^P \left(m^{\alpha}(t)\right)^2
    \label{eq:definition_qm}
\\
    q(t) 
    &= 
        \frac{1}{N}\sum_{\alpha=1}^P
        \sum_{i=1}^{n}
        \left(x^{\alpha}_i(t)\right)^2
    \label{eq:definition_q}
\end{align}
\end{subequations}
are the order parameters that specify, respectively, 
the total activity variance, 
and the variance of mean population activities 
(macroscopic activity). 
In order to analyze the evolution of these order parameters 
we combine the single-neuron evolution equations  
(\ref{eq:update_with_currents}) 
with 
(\ref{eq:current_with_fields}) 
and (\ref{eq:definition_order_parameters}), 
average over quenched disorder, 
and obtain a coupled pair of mean-field equations 
(see Appendix A for the derivation)
\begin{subequations}    
\label{eq:evolution_order_parameters}
\begin{align}
q_m(t+1) 
&= 
\int D\tilde{z} \left[
\int Dz  
\phi\left(
\sigma_{\mu}\sqrt{q_m(t)}\tilde{z} + \sigma\sqrt{q(t)} z
\right)
\right]^2
\label{eq:mf-nd-qm}
\\
q(t+1) 
&= 
\int Dz \left[ 
\phi\left(
\sqrt{\sigma_{\mu}^2 q_m(t)+\sigma^2 q(t)}z 
\right)
\right]^2
\label{eq:mf-nd-q}
\end{align}
\end{subequations}
where $\int Dz f(z) \equiv \int_{-\infty}^{\infty}dz f(z) \exp(-z^2/2)/\sqrt{2\pi}$. 
Fixed points $(q,q_m)$ of the mean-field equations 
(\ref{eq:evolution_order_parameters}) 
specify dynamical regimes (phases) of the network. 

Before we proceed with the analysis of the mean-field equations, 
we note that the mean population activities behave like  
an effective recurrent neural network, 
with units corresponding to populations. 
Indeed, let us combine 
(\ref{eq:update_with_currents}),
(\ref{eq:input_current_detailed}),
and 
(\ref{eq:mean-population-activities}),
take the limit $n \to \infty$, 
and apply the law of large numbers, leading to
\begin{equation}
    m^{\alpha}(t+1)
    =
    \phi_{\sigma\sqrt{q}}\left(
    \mathrm{g}(\sigma \sqrt{q})\sum_{\beta=1}^P \mu^{\alpha\beta} 
    m^{\beta}(t) 
    \right)
    \label{eq:update-ms-effective}
\end{equation}
with 
${
\phi_{a}(\mathrm{g}(a)x) 
     \equiv
     \int Dz 
    \phi\left(
    x + a z
    \right)
}
$
and ${\mathrm{g}(a)\equiv \int\Dz\phi'(a z)}$ chosen such that 
${\phi_a'(0) = \frac{d \phi_a(x)}{dx}|_{x=0}=1}$. 
These considerations lead to a natural decomposition of 
network dynamics into two levels: 
macroscopic dynamics of means (activities of populations)
and microscopic dynamics of neurons within populations around the means. 
A non-zero value of $q_m$ signals that the macroscopic dynamics 
is non-quiescent. 
We will refer to such dynamics as coherent, 
since in this case neural activities within populations are correlated. 
Similarly, a non-zero value of $q$ signals that the overall 
activity of the network is non-quiescent. 
Finally, $q-q_m$ measures the average level of variability 
of neural activity within populations 
(microscopic dynamics).  

In the following we assume that $\phi$ is 
continuous, odd, and non-decreasing with 
$\phi(\infty)=1$ and $\phi'(0)=1$. 
Thus, our results apply to the widely used $\tanh$ activation function, 
although in our computer simulations and some of the mathematical analysis we instead specialize to 
\begin{equation}
\phi_{\infty}(x) 
= 
\erf\left(\frac{\sqrt{\pi}}{2}x\right)   
\label{eq:phi_infinity}
\end{equation}
This implies ${\phi(0)=0}$, 
hence ${q=q_m=0}$ is a fixed point of the mean-field equations 
(\ref{eq:evolution_order_parameters}).  
The linearization around this fixed point  
\begin{subequations}
        \label{eq:mf-order-parameters-linearized}
\begin{align}
    q_m(t+1) 
    &= 
    \sigma_{\mu}^2 q_m(t)
    \label{eq:mf-nd-qm-linearized}
    \\
    q(t+1) 
    &= 
    \sigma_{\mu}^2 q_m(t)+\sigma^2 q(t)
    \label{eq:mf-nd-q-linearized}
    \end{align}
\end{subequations}
demonstrates that 
the quiescent state with $\boldsymbol{x}=0$ 
is stable for $\sigma<1$ and $\sigma_{\mu}<1$. 
Outside of this regime the fixed point with $q=q_m=0$ 
is unstable and a new stable fixed point with $q>0$ signals an active state, 
which can be subdivided into three qualitatively different phases 
and two associated phase transitions. 
The first transition is indicated by the bifurcation 
in the mean-field evolution equation of $q_m$ (\ref{eq:mf-nd-qm}) 
at $\sigma_{\mu}^*$ (for fixed $\sigma$).
Below the transition ($\sigma_{\mu} < \sigma_{\mu}^*$) 
the fixed point is such that $q_m=0$, 
translating into $m^{\alpha}(t)=0$ for all populations. 
In this phase activities within populations are not coherent. 
Above the transition ($\sigma_{\mu} > \sigma_{\mu}^*$) the fixed point with $q_m=0$ 
is unstable and a stable fixed point with $q_m>0$ characterizes the steady state, 
translating into non-zero mean population activities 
(coherent chaos \cite{landau2018coherent}). 
In order to derive an expression for $\sigma_{\mu}^*$, we linearize 
(\ref{eq:evolution_order_parameters}) 
around $q>0$ and $q_m=0$ and find
\begin{equation}
    \sigma_{\mu}^*
    =
    \frac{1}{\mathrm{g}(\sigma\sqrt{q})}
    =
    \frac{1}{\int Dz \phi'(\sigma\sqrt{q}z)}
\end{equation}
Although, as we show below, in this model the active state is necessarily chaotic, 
we can further distinguish two phases that differ 
in the nature and dimensionality of the chaotic activity manifold. 
When the value of $\sigma_{\mu}$ is relatively low, 
the corresponding chaotic attractor 
spans a high dimensional subspace and any random perturbation is rapidly expanded.
In contrast, for $\sigma_{\mu} \gg \sigma$ the chaotic attractor is confined to a relatively 
low dimensional manifold and perturbations in random directions are contracted 
with high probability. 
The directions of expansion are aligned with the macroscopic activity. 

Our derivation is based on the analysis of the evolution 
of an infinitesimal perturbation tracking 
the activity difference between two replicas with shared synaptic weights.  
As in \cite{molgedey1992suppressing}, discrete-time dynamics and self-averaging 
allow us to infer the maximal Lyapunov exponent (MLE) from a single-step 
expansion rate of the perturbation, but here we need to decompose the perturbation into two components:
${
    \epsilon_i^{\alpha}(t) 
    = 
    r^{\alpha}_i(t) + \delta m^{\alpha}(t)
    }
    \label{eq:perturbation-decomposition}
$,
with macroscopic perturbation 
$\delta m^{\alpha}(t)$ lying in a $P$-dimensional 
subspace of the $N$-dimensional neural activity space and 
microscopic perturbation 
$r^{\alpha}_i(t)$ lying in the $(N-P)$-dimensional orthogonal 
complement defined by $P$ constraints 
$\sum_i r_i^{\alpha}(t)=0$. 
The linearization of (\ref{eq:update_with_currents}) 
around $\boldsymbol{\epsilon}(t)=0$ gives
   \begin{equation}
        \epsilon_i^{\alpha}(t+1) 
        =
        \phi'\left(
        \sum_{j, \beta} 
            J_{i j}^{\alpha \beta} 
            x_j^{\beta}(t)
        \right)
        \sum_{j', \beta'}
            J_{i j'}^{\alpha \beta'}
            \epsilon^{\beta'}_{j'}(t)
        \label{eq:evolution_perturbation}
    \end{equation}
    where, as before, 
    $J_{ij}^{\alpha\beta} = 
      \frac{\sigma_{\mu}}{\sqrt{P}}\frac{1}{n}\tilde{z}^{\alpha \beta}  + \frac{\sigma}{\sqrt{N}} z^{\alpha \beta}_{ij}
    $.
    To summarize the evolution of the perturbation within each subspace we define 
    ${q^{\epsilon}(t)=\frac{1}{N}\| \boldsymbol{\epsilon}(t) \|^2}$
    and 
    ${q_m^{\epsilon}(t)=\frac{1}{N}\| \boldsymbol{\delta m}(t) \|^2}$. 
In words, $q^{\epsilon}_m(t)$ and $q^{\epsilon}(t)-q^{\epsilon}_m(t)$ denote 
normalized squared lengths of the perturbation at time $t$ within, 
respectively, macroscopic and microscopic subspaces.
We rewrite (\ref{eq:evolution_perturbation}) as
$
{
        \epsilon_i^{\alpha}(t+1) 
        =
        \eta^{\alpha}_i(t)
        \phi'\left(
          h^{\alpha}_i(t)
        \right)
}
$
where Gaussian fields 
${h^{\alpha}_i(t) = \sigma_{\mu} \sqrt{q_m} \tilde{z}^{\alpha} + \sigma \sqrt{q} z^{\alpha}_i}$ 
and
${\eta^{\alpha}_i(t)=\sigma_{\mu} \sqrt{q_m^{\epsilon}(t)} \tilde{\zeta}^{\alpha}+\sigma \sqrt{q^{\epsilon}(t)} \zeta^{\alpha}_i}$ 
are assumed to be independent 
(i.e., ${z^{\alpha}_i, \tilde{z}^{\alpha}, \zeta^{\alpha}_i, \tilde{\zeta}^{\alpha}}$
are independent standard Gaussian random variables), 
which follows from the observation that 
$\boldsymbol{x}(t)$ and $\boldsymbol{\epsilon}(t)$ are generically expected to be orthogonal. 
Finally, we average over the Gaussian fields, leading to 
(see Appendix B for the derivation)
 \begin{equation}
    \begin{pmatrix}
        q_m^{\epsilon}(t+1)
        \\
        q^{\epsilon}(t+1)
        -
        q_m^{\epsilon}(t+1)
    \end{pmatrix} 
    =
    \boldsymbol{D} 
    \begin{pmatrix}
        q_m^{\epsilon}(t)
        \\
        q^{\epsilon}(t)
        -
        q_m^{\epsilon}(t)
    \end{pmatrix}
\end{equation}
with 
\begin{equation}
    \boldsymbol{D} 
    =
     \begin{pmatrix}
        R^2_{coherent} & 0
        \\
        C & R^2_{random}
    \end{pmatrix}
\end{equation}
where 
$C =(\sigma_{\mu}^2/\sigma^2 + 1) R^2_{random} - R^2_{coherent} \geq 0$
and 
\begin{subequations}
\begin{align}
    R^2_{coherent}
    &= 
    \sigma_{\mu}^2 
    \int D\tilde{z}
    \left[ 
    \int Dz
    \phi'(
      \sigma_{\mu} \sqrt{q_m} \tilde{z}
      +  
      \sigma \sqrt{q} z
    )
    \right]^2
    \\
        R^2_{random}
    &= 
    \sigma^2 
    \int Dz 
    \left[ 
      \phi'
      \left(
      \sqrt{\sigma_{\mu}^2 q_m
      +  
      \sigma^2 q }
      z
      \right)
      \right]^2
\end{align}
\end{subequations}
are the eigenvalues of the $\boldsymbol{D}$. 
The long-term evolution of a generic perturbation will be dominated 
by the larger of the two. Therefore, the MLE can be calculated as 
${\lambda_{max} = \max(\lambda_{random},\lambda_{coherent})}$ 
where 
$\lambda_X = \ln(R_X)$.

The interactions between macroscopic and microscopic 
dynamics are asymmetric. 
Although a macroscopic perturbation has a direct effect on microscopic activity, 
a microscopic perturbation does not transmit to macroscopic activity due to averaging. 
This translates into the lower-triangular form of $\boldsymbol{D}$. 
Moreover, high-dimensional chaos resulting from chaotic microscopic 
dynamics can be suppressed by high levels of 
macroscopic activity. 
Indeed, the asymptotic behavior of $R^2_{random}$ for ${\sigma_{\mu}\gg1}$ reads ${R^2_{random} \approx 
    C_0 
    \sigma^2/\sqrt{\sigma_{\mu}^2 q_m + \sigma^2 q}}$.
Since $q_m$ and $q$ are non-decreasing and bounded functions of $\sigma_{\mu}$, 
for any fixed $\sigma$ there exists $\sigma_{\mu}^{EoHC}$ 
such that for $\sigma_{\mu}>\sigma_{\mu}^{EoHC}$ 
a random perturbation within the $(N-P)$-dimensional 
microscopic subspace is contracted ($R^2_{random} < 1$), 
i.e. high-dimensional chaos is suppressed. 
Based on this analysis, we can expect that around $\sigma_{\mu}^{EoHC}$ 
the system will transition between low-dimensional and high-dimensional chaos. 
\begin{figure}
        \includegraphics[width=0.49\linewidth]{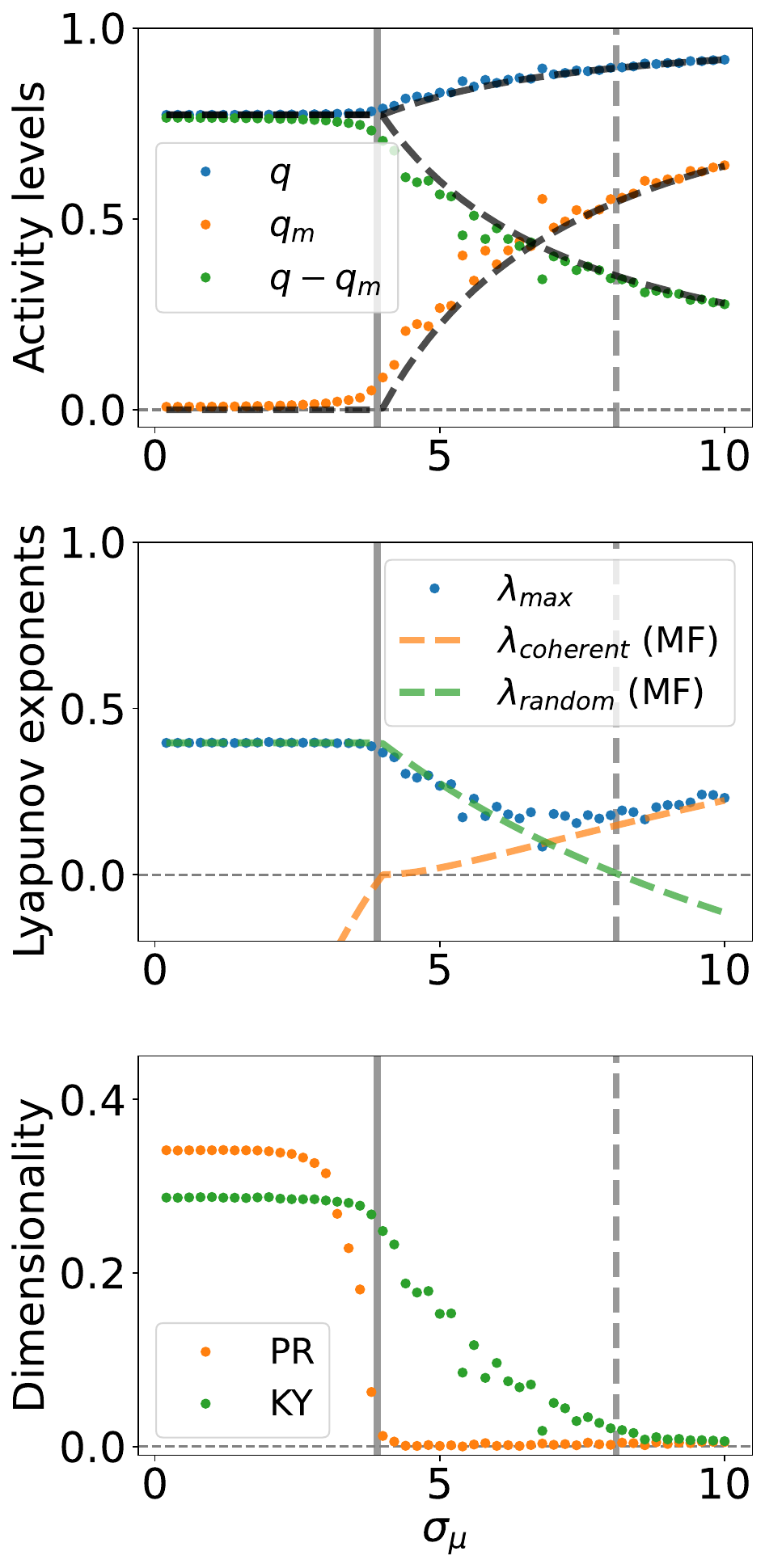}
        \includegraphics[width=0.49\linewidth]{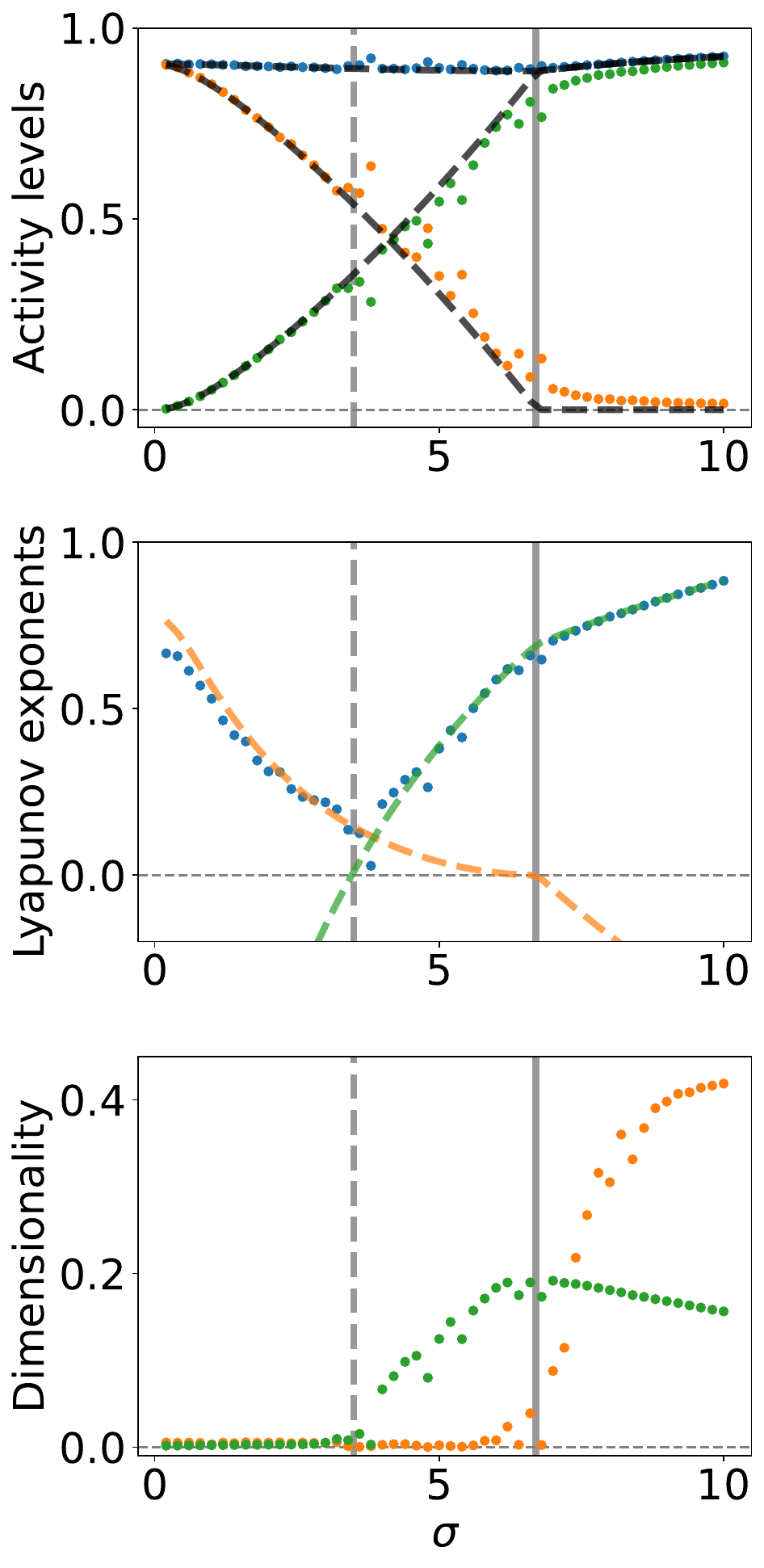}
    \caption{
    Statistics of the typical 
    autonomous dynamics as functions of 
    $\sigma_{\mu}$ for $\sigma=3.5$
    \textit{(left)} 
    and as functions of
    $\sigma$ for $\sigma_{\mu}=8$ 
    \textit{(right)}. 
    Dots and dashed lines correspond to results of computer simulations
    and predictions of the mean-field theory, respectively. 
    Vertical gray lines represent 
    transition lines: 
    $\lambda_{coherent}=0$ (solid) 
    and 
    $\lambda_{random}=0$ (dashed). 
    \textit{Top:} 
    Total ($q$), 
    macroscopic ($q_m$), 
    and microscopic ($q-q_m$) variances. 
    \textit{Center:} 
    Maximal Lyapunov exponents computed either 
    directly from simulations or 
    as within-subspace theoretical predictions. 
    \textit{Bottom:}
    Normalized dimensionality 
    of the steady-state neural activity 
    manifold as measured by the 
    covariance matrix (PR dimension) 
    or Lyapunov exponents (KY dimension).
    }
    \label{fig:just_simulations}
\end{figure}

\textit{Computer simulations.---}We test this prediction with computer simulations 
(see Appendix E for details). 
In Fig.~\ref{fig:eigvals_and_activities}, 
we present eigenvalues of the weight matrix and example 
activities of neurons and populations in 
the four phases predicted by the mean-field theory. 
In Fig.~\ref{fig:just_simulations}, 
we illustrate relevant statistics of the dynamics as functions of control parameters. 
As expected, with fixed $\sigma>1$ and increasing $\sigma_{\mu}$, 
we observe a region with a substantial drop in the values of the 
Kaplan-Yorke (KY or Lyapunov) dimension \cite{kaplan1979chaotic,frederickson1983liapunov,ott2002chaos}, 
where the Lyapunov dimension transitions from being extensive in $N$ 
to being extensive in $P$. 
In contrast, the participation ratio (PR) dimension 
\cite{rajan2010inferring,gao2017theory,recanatesi2019dimensionality,recanatesi2022scale,clark2023dimension}
exhibits an abrupt transition around 
$\sigma_{\mu}^*$, 
leading to a region where these two measures of attractor dimensionality 
differ by orders of magnitude. 
A similar region is observed with fixed $\sigma_{\mu}>1$ and increasing $\sigma$. Additionally, in this case the Lyapunov dimension features a peak and slowly decreases at large values of $\sigma$. This effect has been previously noted and explained in the single-population model \cite{engelken2023lyapunov}. 
Other quantitative predictions of our mean-field analysis are also confirmed by the simulations. 
In particular, the MLE admits a minimum at intermediate values 
of $\sigma$ (for fixed $\sigma_{\mu}$) 
or $\sigma_{\mu}$ (for fixed $\sigma$),
suggestive of an effective competition between 
macroscopic and microscopic dynamics. 
In other words, both adding modular structure 
(increasing $\sigma_{\mu}$) 
to mostly random connectivity 
(large initial $\sigma$ and small initial $\sigma_{\mu}$)
and 
injecting noise (increasing $\sigma$) 
to strongly modular connectivity (large initial $\sigma_{\mu}$ and small $\sigma$) 
lead to attenuation of chaos.
These results are summarized as a phase diagram in Fig.~\ref{fig:phase_diagram}. 
Furthermore, our preliminary simulations 
confirm that the same qualitative 
results are obtained in the continuous-time 
version of the model 
\footnote{To be published elsewhere}. 

\textit{Multilevel generalization.---}In biological neural circuits we expect each population to consist of multiple sub-populations, 
which in turn may be further subdivided into sub-sub-populations, and so on. 
To capture such a hierarchical connectivity organization we introduce a generalization of our model in the following, recursive manner:
    \begin{equation}
        \boldsymbol{J^{[i]}} = 
        \boldsymbol{J^{[i-1]}} \otimes \boldsymbol{O^{(P_i)}} + \sigma_i \boldsymbol{\Xi^{(N_i)}}
        \label{eq:recursive_multilevel_connectivity}
    \end{equation}
where $N_i = \Pi_{j=0}^i P_j$, 
$P_0 = 1$, $\boldsymbol{J^{[0]}} = 0$, 
and $\sigma_i$ and $P_i$ are parameters 
specific to level $i$. 
Our basic model is equivalent to 
the case with two levels 
($\boldsymbol{J}=\boldsymbol{J^{[2]}}$) and 
${\sigma_1 = \sigma_{\mu}}$, 
${\sigma_2 = \sigma}$,
${P_1 = P}$, ${P_2 = n}$. 

To facilitate the analysis we make use of 
tensor notation with indices that specify 
the unit at each level of the hierarchy. 
We rewrite the evolution equation as
    ${x_{\alpha_1\dots\alpha_L}(t+1)
    =
    \phi\left(
    h_{\alpha_1\dots\alpha_L}(t)
    \right)
    }$
    with 
    ${
     h_{\alpha_1\dots\alpha_L}(t) 
     = 
     \sum_{\beta_1,\dots,\beta_L}
     J_{\alpha_1\dots\alpha_L}^{\beta_1\dots\beta_L}
     x_{\beta_1\dots\beta_L}(t)
     }$, 
and introduce mean activities at level $j$ as
    ${m_{\alpha_1\dots\alpha_j}(t) 
    =
    (N_j/N_L)
    \sum_{\alpha_{j+1},\dots,\alpha_L} 
    x_{\alpha_1\dots\alpha_L}(t)}$. 
Since we only make use of sums over adjacent right-most indices, 
this simplified notation does not introduce any ambiguity.
We assume $\forall_i P_i \gg 1$ and substitute 
    ${  
    \sum_{j=1}^L
    \sigma_j \sqrt{q_j(t)}
    z^{[j]}_{\alpha_1\alpha_2\dots\alpha_j}}$ 
for the preactivations ${h_{\alpha_1\dots\alpha_L}(t)}$, 
where $z^{[j]}_{\alpha_1\dots\alpha_j}$ 
are i.i.d. standard normal random variables 
and 
    ${q_j(t)
    = 
    \frac{1}{N_j}
    \sum_{\alpha_1, \dots, \alpha_j }
    \left( m_{\alpha_1 \dots \alpha_j}(t) \right)^2}$
are the order parameters that in effect evolve as 
(see Appendix A for the derivation)
\begin{equation}
\label{eq:dmf-qs-general}
    q_j(t+1)
    =
    \int D\tilde{z}
    \left(
    \int Dz
    \phi\left(
    \sqrt{A_j(t)}
    \tilde{z}
    +
    \sqrt{A_L(t) - A_j(t)} 
    z
    \right)
    \right)^2
\end{equation}
where $A_j(t)=\sum_{i=1}^j \sigma_i^2 q_i(t)$. 
As in the two-level case, here also, each level contributes its own (maximal within the relevant subspace) Lyapunov exponent 
$\lambda_j = \ln R_i$, 
which in the mean-field limit can be expressed as 
(see Appendix B for the derivation)
\begin{equation}
R^2_{j}
    =
    \sigma_j^2
    \int D\tilde{z}
    \left(
    \int Dz
    \phi'\left(
    \sqrt{A_j} 
    \tilde{z}
    +
    \sqrt{A_L - A_j} 
    z
    \right)
    \right)^2
\label{eq:dmf-Rs-general}
\end{equation}
where 
$A_j=\lim_{t\to\infty}A_j(t)$.
In the special case of 
${\phi(x) = \phi_{\infty}(x)}$ 
we evaluate all relevant integrals and arrive at 
closed-form expressions
\begin{equation}
    \sigma^2_j 
    =
    \frac{2}{\pi q_j}
    \frac{\sin\left(\frac{\pi}{2}q_j\right)-\sin\left(\frac{\pi}{2}q_{j-1}\right)}{1 - \sin\left(\frac{\pi}{2}q_L\right)}
    \label{eq:multilevel-erf-sigma}
\end{equation}
and
\begin{equation}
    R_j^2
    =
    \frac{2}{\pi q_j}
    \frac{\sin\left(\frac{\pi}{2}q_j\right)-\sin\left(\frac{\pi}{2}q_{j-1}\right)}{ \cos\left(\frac{\pi}{2}q_j\right)}
    \label{eq:multilevel-erf-R2}
\end{equation}
Eq.~(\ref{eq:multilevel-erf-sigma}) specifies a sequence 
of control parameters that gives rise to 
a steady state described by an admissible 
sequence of order parameters 
${0=q_0 \leq q_1 \leq q_2 \leq \dots \leq q_L < 1}$
\footnote{
This sequence is not unique when some of the highest-level activities are set to zero, i.e.  
${q_1=q_2=\dots=q_j=0}$. 
In this case, the corresponding limiting values of 
${\sigma_1=\sigma_2=\dots=\sigma_j}$ 
specify the maximum values of these parameters (phase boundary), 
as any subsequence $(\sigma_i)_{i=1}^j$ within the $j$-dimensional hypercube leads to the same values of ${(q_i)_{i=1}^L}$.
}. 

Given the explicit form of (\ref{eq:multilevel-erf-R2}), 
we are ready to investigate how the MLE 
depends on the levels of network activity. 
Let $k$ be such that $q_{k-1}=0$ and $q_k>0$. 
The corresponding MLE
$\lambda_k = \frac{1}{2}\ln\left(\tan\left(\frac{\pi}{2}q_k\right)/\left(\frac{\pi}{2} q_k\right)\right)$ 
is positive and does not depend on the 
order parameters at lower levels. 
Thus, the non-quiescent steady state of this model  
is always chaotic. 
However, in deep hierarchical structures with $L\gg1$ we can expect the overall 
MLE to be close to zero, 
as long as many levels are active and approximately balanced, i.e., if 
${\Delta_j=q_j-q_{j-1}}$ are all of the order of $q_L/L$. 
For $L\gg 1$ we then have ${\Delta_j \ll 1}$ 
and we can expand the level-specific 
MLEs as 
${\lambda_k \approx \pi^2 \Delta_k^2/24}$ 
and, for $j>k$,  
${\lambda_j \approx \frac{1}{2}\ln\left( \frac{\Delta_j}{q_j} + \frac{\pi \tan\left(\frac{\pi}{2}q_j\right)}{4 q_j} \Delta_j^2\right)}$ 
\footnote{In the latter case we retain two terms since the quadratic term 
may dominate for large $q_j$ and finite $L$.}. 
This demonstrates that $\lambda_{max} \approx 0$ for 
$L \gg 1$ and $L^2 \gtrapprox \tan(\frac{\pi}{2}q_L)$. 
Therefore, any mechanism that maintains a loosely balanced 
hierarchical activity of the network 
(i.e., with all $\Delta_j$ of the order of 1/L) 
will also keep it close to the edge of chaos. 
To illustrate this, we devise and test a proof-of-concept adaption algorithm 
\footnote{See Supplementary Material for details}. 
Importantly, although our analysis above assumes 
${\phi(x)=\phi_{\infty}(x)}$, 
we expect its conclusions to be more general since,  
due to an effective coarse-graining, 
the dynamics at higher levels of the hierarchy 
is universal for a wide range of activation functions 
(see Appendix D for details). 

\textit{Conclusion.---}We introduced a modular neural 
network model and studied its autonomous dynamics. 
We showed that, with increasing correlations between within-population weights, 
the suppression of high-dimensional chaotic dynamics proceeds in two steps. 
First, starting from $\sigma_{\mu}^*$ the macroscopic activity of 
populations becomes coherent, 
as signaled by non-zero $q_m$, positive $\lambda_{coherent}$, 
and a low value of the participation ratio dimension. 
At $\sigma_{\mu}^*$, critical macroscopic dynamics coexist 
with non-critical, chaotic microscopic dynamics, 
analogous to the low-dimensional criticality embedded in 
high-dimensional neural dynamics 
recently uncovered in the motor cortex of awake, 
behaving mice \cite{fontenele2024low}. 
Second, high-dimensional chaos is completely suppressed above 
$\sigma_{\mu}^{EoHC}$, 
as signaled by negative $\lambda_{random}$ and a low value of the Lyapunov dimension. 
Moreover, the maximal Lyapunov exponent features a prominent dip at intermediate values of 
weights correlations, indicating that modular synaptic connectivity supports 
dynamics in the vicinity of the edge of chaos and suggesting 
an interesting computational regime that maintains the balance 
between macroscopic and microscopic dynamics. 
This effect is even more pronounced in the multilevel generalization of the model. 
We hypothesize that it provides an example of a general mechanism 
that underpins a potentially ubiquitous phenomenon 
of complex biological systems evolving towards the edge of chaos 
through a loose balance of multiscale, hierarchical dynamics. 
Furthermore, in the brain activities at different levels of hierarchy 
may encode and process distinct information, 
with level-specific coherent-quiescent transitions providing 
a substrate for flexible modulation of information flow. 
\bibliography{references}

\onecolumngrid
\section*{End matter}
\twocolumngrid
\subsection{Appendix A: Derivation of (\ref{eq:dmf-qs-general})} \label{sec:appendixA}
As noted in the main text, we rewrite the evolution 
equation as
\begin{equation}    
    x_{\alpha_1\dots\alpha_L}(t+1)
    =
    \phi\left(
    \sum_{j=1}^L
    \sigma_j \sqrt{q_j(t)}
    z^{[j]}_{\alpha_1\dots\alpha_j}
    \right)
    \label{eq:app_evolution_x_multilevel}
\end{equation}
where $z^{[j]}_{\alpha_1\alpha_2\dots\alpha_j}\sim \mathcal{N}(0,1)$ are i.i.d. 
This decomposition originates from the hierarchical structure of the weight matrix, 
which contains a sum over $L$ levels, 
each contributing an independent source of variability:
\begin{equation}
     J^{\beta_1\dots\beta_L}_{\alpha_1\dots\alpha_L} 
     =
     \sum_{j=1}^L
     \frac{1}{P_{j+1}\dots P_L}\frac{\sigma_j}{\sqrt{N_j}}
     \tilde{z}^{[j]}_{\alpha_1\dots\alpha_j \beta_1\dots\beta_j}
\end{equation}
where 
$\tilde{z}^{[j]}_{\alpha_1\dots\alpha_j \beta_1\dots\beta_j}\sim \mathcal{N}(0,1)$ 
are all independent. 
Equation (\ref{eq:app_evolution_x_multilevel}) follows immediately from 
\begin{equation}
    \sum_{\beta_1\dots \beta_L} 
    \frac{1}{P_{j+1}\dots P_L}\frac{1}{\sqrt{N_j}}
     \tilde{z}^{[j]}_{\alpha_1\dots\alpha_j \beta_1\dots\beta_j}
    x_{\beta_1\dots \beta_L}(t)
    =
    \sqrt{q_j(t)}
    z^{[j]}_{\alpha_1\dots\alpha_j}
\end{equation}
Our goal is to find an evolution equation of
\begin{equation}
    q_j(t)
    = 
    \frac{1}{P_1 \dots P_j}
    \sum_{\alpha_1, \dots, \alpha_j }
    \left( m_{\alpha_1 \dots \alpha_j}(t) \right)^2
    \label{eq:app_def_q}
\end{equation}
where 
\begin{equation}
    m_{\alpha_1 \dots \alpha_j}(t)
    =
    \frac{1}{P_{j+1} \dots P_L}
    \sum_{\alpha_{j+1},\dots,\alpha_L}
    x_{\alpha_1 \dots \alpha_L}(t)
    \label{eq:app_def_m}
\end{equation}
We start by rewritting (\ref{eq:app_def_m}) at time $t+1$ 
using (\ref{eq:app_evolution_x_multilevel})
\begin{equation}
    m_{\alpha_1 \dots \alpha_j}(t+1)
    =
    \frac{1}{P_{j+1}\dots P_L}
    \sum_{\alpha_{j+1},\dots,\alpha_L}
     \phi\left(
    \sum_{i=1}^L
    \sigma_i \sqrt{q_i(t)}
    z^{[i]}_{\alpha_1\dots\alpha_i}
    \right)
    \label{eq:app_m_intermediate}
\end{equation}
We note that all terms under the activation function 
with $i > j$ contain at least a single index $\alpha_k$ 
that is averaged over, which in the limit of 
$P_k \to \infty$ converges to the expectation over the corresponding standard normal variable. 
In contrast, terms with $i \leq j$ do not contain  
summation indices $\alpha_{j+1},\dots, \alpha_L$ and 
are not affected by the averaging. 
Thus, the right-hand side (RHS) of (\ref{eq:app_m_intermediate}) 
can be rewritten as
\begin{equation}
    \int_{\mathbb{R}^{L-j}} \Dz_{j+1} \dots \Dz_L
    \phi\left(
    \sum_{i=1}^j
    \sigma_i \sqrt{q_i(t)}
    z^{[i]}_{\alpha_1\dots\alpha_i}
    +
    \sum_{k=j+1}^L \sigma_k \sqrt{q_k(t)} z_k 
    \right)
\end{equation}
where $\Dz_i=dz_i \exp\left(-z_i^2/2\right)/\sqrt{2\pi}$.
Due to the stability of the Gaussian distribution 
(convolutions of Gaussian distributions are still Gaussian), 
we can replace the expectation of a function of a sum of Gaussian random variables with an expectation of the same function over a properly scaled single Gaussian random variable, i.e.
\begin{equation}
    \int_{\mathbb{R}^n} Dz_1 \dots \Dz_n 
    f\left(\sum_{i=1}^n a_i z_i\right)
    =
    \int_{\mathbb{R}} Dz 
    f\left(z \sqrt{\sum_{i=1}^n a^2_i}\right),
\end{equation}
reducing (\ref{eq:app_m_intermediate}) to 
\begin{equation}
    m_{\alpha_1 \dots \alpha_j}(t+1)
    =
    \int_{\mathbb{R}} \Dz 
    \phi\left(
    \sum_{i=1}^j
    \sigma_i \sqrt{q_i(t)}
    z^{[i]}_{\alpha_1\dots\alpha_i}
    +
    \sqrt{A_L(t) - A_j(t)} z 
    \right)
    \label{eq:app_m_mf}
\end{equation}
where $A_k(t)=\sum_{i=1}^k \sigma_i^2 q_i(t)$.
In the final step we plug (\ref{eq:app_m_mf}) into 
(\ref{eq:app_def_q}), repeat the process of replacing 
averages with expected values, and subsequently replacing 
expectations over multiple Gaussian random variables with 
an expected value over a single Gaussian random variable.


\subsection{Appendix B: Derivation of (\ref{eq:dmf-Rs-general})}
Similarly to the basic two-level model, we introduce an infinitesimal perturbation 
$\epsilon_{\alpha_1\dots\alpha_L}(t)$ 
and linearize the evolution equation (\ref{eq:update_with_currents}), arriving at a linear 
evolution equation for the perturbation
\begin{equation}
    \epsilon_{\alpha_1\dots\alpha_L}(t+1)
    = 
    \phi'\left(
    \sum_{\beta_1,\dots,\beta_L}
    J^{\beta_1\dots\beta_L}_{\alpha_1\dots\alpha_L} x_{\beta_1\dots\beta_L}(t)
    \right)
    \sum_{\gamma_1,\dots,\gamma_L}
    J^{\gamma_1\dots\gamma_L}_{\alpha_1\dots\alpha_L} \epsilon_{\gamma_1\dots\gamma_L}(t)
    \label{eq:app_evolution_eps_multilevel}
\end{equation}
In analogy to the order parameters used to describe network activity, we introduce 
\begin{equation}
    m^{\epsilon}_{\alpha_1\dots\alpha_j}(t) 
    =
    \frac{1}{P_{j+1}\dots P_L}
    \sum_{\alpha_{j+1},\dots,\alpha_L}
    \epsilon_{\alpha_1\dots\alpha_L}(t)
    \label{eq:app_def_m_eps}
\end{equation}
and
\begin{equation}
    q^{\epsilon}_j(t)
    = 
    \frac{1}{P_1 \dots P_j}
    \sum_{\alpha_1, \dots, \alpha_j }
    \left( m^{\epsilon}_{\alpha_1 \dots \alpha_j}(t) \right)^2
    \label{eq:app_def_q_eps}.
\end{equation}
We replace both sums in 
(\ref{eq:app_evolution_eps_multilevel})
with Gaussian fields 
\begin{equation}
    \epsilon_{\alpha_1\dots\alpha_L}(t+1)
    =
    \phi'\left(
    \sum_{j=1}^L 
    \sigma_j \sqrt{q_j(t)} 
    z^{[j]}_{\alpha_1\dots\alpha_j}
    \right)
    \sum_{k=1}^L
    \sigma_k \sqrt{q^{\epsilon}_k(t)} 
    \tilde{z}^{[k]}_{\alpha_1\dots\alpha_k}. 
\end{equation}
Crucially, we assume that 
$z^{[j]}_{\alpha_1\dots\alpha_j}$ 
and
$\tilde{z}^{[j]}_{\alpha_1\dots\alpha_j}$ 
are independent, which follows from the assumption 
that $\mathbf{x}(t)$ and $\boldsymbol{\epsilon}(t)$ 
are generically orthogonal within any of the considered subspaces (at each level of coarse-graining). 
Given this assumption, we can replace averages with expectations over independent Gaussian random variables. 
After straightforward manipulations we arrive at
\begin{equation}
    q_j^{\epsilon}(t+1)
    =
    A_j^{\epsilon}(t)
    \int\limits_{-\infty}^{\infty}
    D\tilde{z}
    \left(
    \int\limits_{-\infty}^{\infty}
    Dz
    \phi'\left(
    \sqrt{A_j(t)}\tilde{z}
    +
    \sqrt{A_L(t) - A_j(t)} z
    \right)
    \right)^2
    \label{eq:app_q_eps_ev}
\end{equation}
where $A_j^{\epsilon}(t)=\sum_{i=1}^j \sigma_i^2 q^{\epsilon}_i(t)$ 
comes from the term 
$\sum_{k=1}^L \sigma_k \sqrt{q_k^{\epsilon}(t)}\tilde{z}^{[k]}_{\alpha_1\dots\alpha_k}$.
In the steady-state, ${q_j=\lim_{t\to\infty}q_j(t)}$ 
and 
${A_j=\lim_{t\to\infty}A_j(t)}$ 
are constant and (\ref{eq:app_q_eps_ev}) is linear dynamical system of the form ${\mathbf{q^{\epsilon}}(t+1) = \mathbf{D} \mathbf{q^{\epsilon}}(t)}$, where $\mathbf{D}$ is an $L\times L$ time-independent triangular matrix. Its diagonal entries correspond exactly to its eigenvalues and are given by (\ref{eq:dmf-Rs-general}).


\subsection{Appendix C: Relevant integrals}
In order to derive (\ref{eq:multilevel-erf-sigma}) and (\ref{eq:multilevel-erf-R2}) from (\ref{eq:dmf-qs-general}) and (\ref{eq:dmf-Rs-general}) we used the following identities:
\begin{equation}
    \int_{\mathbb{R}} \Dz \phi_{\infty}(a z + b) 
    =
    \phi_{\infty}\left( 
        \frac{b}{\sqrt{1 + \pi a^2/2 }}
    \right)
    \label{eq:erf_id_linear}
\end{equation}
\begin{equation}
    \int_{\mathbb{R}} \Dz \left( \phi_{\infty}(c z) \right)^2 
    =
    \frac{4}{\pi}\arctan\left(
    \sqrt{1 + \pi c^2}
    \right) 
    - 
    1
    \label{eq:erf_id_square}
\end{equation} 
To prove (\ref{eq:erf_id_linear}), we introduce 
$f(a,b)=\int_{\mathbb{R}} \Dz \phi_{\infty}(a z + b)$ 
and calculate $\partial_b f(a,b)$. 
Since $\phi_{\infty}'(x) = \exp(-\pi x^2/4)$, 
this results in a Gaussian integral (over the entire real line) which is straightforward to evaluate, leading to 
\begin{equation}
\partial_b f(a,b) = 
\frac{1}{\sqrt{1 + \frac{\pi}{2} a^2} }
\exp\left(
-\frac{\frac{\pi}{4}b^2}{1 + \frac{\pi}{2}a^2}
\right)
\end{equation}
Thus, $f(a,b)$ is given by the indefinite Gaussian integral
\begin{equation}
    f(a,b) 
    =
    \phi_{\infty}\left( 
    \frac{b}{\sqrt{1+\pi a^2/2}}
    \right)
    +
    C(a)
\end{equation}
where $C(a)$ is an unknown function of $a$. 
Finally, due to the fact that $\phi_{\infty}$ 
is odd we find that $C(a)=f{(a,0)=0}$.$\qed$

To prove (\ref{eq:erf_id_square}), we introduce 
$g(c)=\int \Dz \left[\phi_{\infty}(c z)\right]^2$ and calculate its derivative
\begin{equation}
    g'(c) 
    =
    \sqrt{\frac{2}{\pi}}
    \int\limits_{-\infty}^{\infty} dz 
    \exp\left( 
    - \frac{z^2}{2} \left(1 + \frac{\pi}{2} c^2\right)
    \right) 
    z 
    \phi_{\infty}(c z)
    \label{eq:g_prime_intermediate}
\end{equation}
We rewrite the RHS of (\ref{eq:g_prime_intermediate}) as
\begin{equation}
    \sqrt{\frac{2}{\pi}}
    \int\limits_{-\infty}^{\infty} dz 
    \frac{d}{dz}
    \left[-
    \frac{1}{1+\frac{\pi}{2}c^2}
    \exp\left( 
    - \frac{z^2}{2} \left(1 + \frac{\pi}{2} c^2\right)
    \right) 
    \right] 
    \phi_{\infty}(c z)
\end{equation}
and, applying integration by parts, arrive at
\begin{equation}
    g'(c)
    =
    \frac{2c}{\left(1 + \frac{\pi}{2}c^2\right) \sqrt{1 + \pi c^2}}
\end{equation}
Next, we set $u=\sqrt{1 + \pi c^2}$ and perform 
integration by substitution, which gives
\begin{equation}
    g(c) 
    =
    \frac{4}{\pi}\arctan\left(
    \sqrt{1 + \pi c^2}
    \right)
    + 
    C_0
\end{equation}
To find the value of $C_0$, we observe that 
$g(0)=0$ (since $\phi_{\infty}(0)=0$), 
from which we infer $C_0=-1$.
$\qed$


\subsection{Appendix D: Universality} 
Assume $\phi$ to be odd with $\phi(\infty)=1$ 
and a single-peaked $\phi'$. 
If most lower levels contribute to the activity of the multilevel system, 
we have $\sigma_i>1$ for most levels. 
Thus, at high-enough levels of the hierarchy, 
$j\ll L$, we have $A_L - A_j \gg 1$ and the limit 
$\lim_{a\to\infty} a \phi'(a x) = 2 \delta(x)$ 
provides a way of approximating 
(\ref{eq:dmf-qs-general}) and (\ref{eq:dmf-Rs-general}) as 
\begin{equation}
    {q_j(t+1) 
    \approx 
    \int Dz
    \left(
    \phi_{\infty}
    \left(
    B_j(t) z
    \right)
    \right)^2}
\end{equation}
and
\begin{equation}
   {R_j^2
    \approx
    \tilde{\sigma}_j^2
    \int Dz
    \left(
    \phi'_{\infty}
    \left(
    B_j z
    \right)
    \right)^2}, 
\end{equation}
where ${\tilde{\sigma}_j^2 = \frac{2}{\pi} \frac{\sigma_j^2}{A_L - A_j}}$, 
${B_j(t) = \sqrt{\frac{2}{\pi}}\sqrt{\frac{A_j(t)}{A_L(t)-A_j(t)}}}$, 
${B_j=\lim_{t\to\infty}B_j(t)}$, 
and 
${\phi_{\infty}(x) = \lim_{a\to\infty}\phi_a(x)}$ 
is given by (\ref{eq:phi_infinity}). 
Due to (\ref{eq:erf_id_linear}), the two formulas above match the mean-field 
equations for a multilevel system with the activation function set to $\phi_{\infty}$. 
Indeed, direct computations similar to (\ref{eq:update-ms-effective})
confirm that $\phi_{\infty}$ is the activation function 
of the effective evolution of the higher level activities. 
In other words, $\phi_{\infty}$ 
is a fixed point of the coarse-graining procedure for a wide range of 
underlying microscopic activation functions. 
This illustrates the special status of 
$\phi_{\infty}$ as the effective activation function of 
higher, population level activities, 
and it provides a justification for the expectation 
that our results extend with no qualitative changes 
to many activation functions. 

\subsection{Appendix E: Details of the computer simulations}
Fig.~\ref{fig:eigvals_and_activities}: 
For visualization purposes, 
network sizes differed between 
weight visualizations (${n=8,P=8}$),
eigenvalues panels (${n=5,P=200}$), 
and the simulations (${n=200,P=100}$). 
Networks were simulated for $220$ steps starting from random initial conditions, 
with activities from the last $20$ steps used for plotting. 
The random seed was shared across columns. 
In Fig.~\ref{fig:just_simulations}, 
recurrent neural networks with $P=n=100$ were simulated for $22000$ steps 
starting from random initial conditions 
(i.i.d. ${x^{\alpha}_i\sim\mathcal{N}(0,1)}$). 
Weights and initial conditions were reinitialized for each pair of values of control parameters 
$(\sigma, \sigma_{\mu})$. 
In order to ensure that the statistics were collected close to the steady state, 
first $2000$ steps were discarded. 
Network activity at all remaining $20000$ steps
were used to calculate $q$, $q_m$, 
and the participation ratio dimension. 
Network activity during the last $500$ steps were used to compute 
Lyapunov exponents and the associated Lyapunov dimension. 
Dimensionalities were normalized by the total number of neurons ($10^4$). 
For numerical stability, QR decomposition was used in computations of 
the Lyapunov exponents; for details of the algorithm see 
\cite{vogt2022lyapunov}. 
The code, available online at 
\url{https://github.com/AllenInstitute/modular-networks-theory}, 
was written in Python \cite{python} utilizing JAX \cite{jax2018github} for GPU acceleration. 
\clearpage
\title{Supplementary Material: Hierarchy of chaotic dynamics in random modular networks}

\maketitle

\onecolumngrid

\section{$\sigma(\Delta)$ adaptation algorithm}
\label{app:adaptation}
In the main text, we argue that an adaptation algorithm driving the system toward a balance 
across hierarchical activity levels naturally positions it near the edge of chaos. 
To support this claim, here we introduce and test a simple adaptation algorithm. 
It is important to note that the algorithm presented here serves as a proof of concept 
rather than a specific model that we propose the brain, or any natural system, implements.

The algorithm is based on the observation that 
a hierarchical modular system with $L$ levels should be close 
to the edge of chaos as long as ${\Delta_i=q_i-q_{i-1}}$ are 
all of the order of $q_L/L$, 
where $q_L$ is the total activity variance of the network.
Thus, given a desired total level of activity $\hat{q}_L$, 
we set the desired within-level activities to
\begin{equation}
\hat{\Delta}_i=\frac{\hat{q}_L}{L}
\end{equation} 
We assume that the synaptic weights are drawn randomly according to the 
multilevel version of our connectivity model (18), but with a time-dependent 
level-specific control parameters $\sigma_i(t)$, i.e.
\begin{equation}
    \boldsymbol{J^{[i]}}(t) = 
    \boldsymbol{J^{[i-1]}}(t) \otimes \boldsymbol{O^{(P_i)}} + \sigma_i(t) \boldsymbol{\Xi^{(N_i)}}
    \label{eq:SM_recursive_multilevel_connectivity}
\end{equation}
For simplicity, the individual entries $\boldsymbol{\Xi^{(N_i)}}$ are fixed. 
At each time step of the adaptation process, $\Delta_i(t)$ are computed based on the network activity $\boldsymbol{x}(t)$, and $\sigma_i(t)$ are updated according to
\begin{equation}
    \sigma_i(t+1)
    =
    \sigma_i(t)
    +
    \eta 
    \left(
    \hat{\Delta}_i
    -
    \Delta_i(t) 
    \right)
    \label{eq:SM_adaptation}
\end{equation}
where $\eta$ is a fixed learning rate. 
Equation (\ref{eq:SM_adaptation}) was not derived from any 
cost function. Instead, it is based on the intuitive observation 
that, all else being equal, 
$\Delta_i$ is expected to be a non-decreasing function of $\sigma_i$. 
Note also that our choice of $\hat{\Delta}_i$ is somewhat 
arbitrary and is not actually optimal, 
i.e. it does not minimize the maximal Lyapunov exponent for a given $\hat{q}_L$. 
Nonetheless, as we argue in the main text, it should still 
lead to low (and positive) values of the maximal 
Lyapunov exponent if $L$ is large enough. 
Our numerical experiments, described in detail below, show that this 
is already true even for $L=2$ and $L=3$, 
as long as $q_L$ is not too close to $1$. 

\begin{figure}
    \centering
    \includegraphics[width=0.195\linewidth]{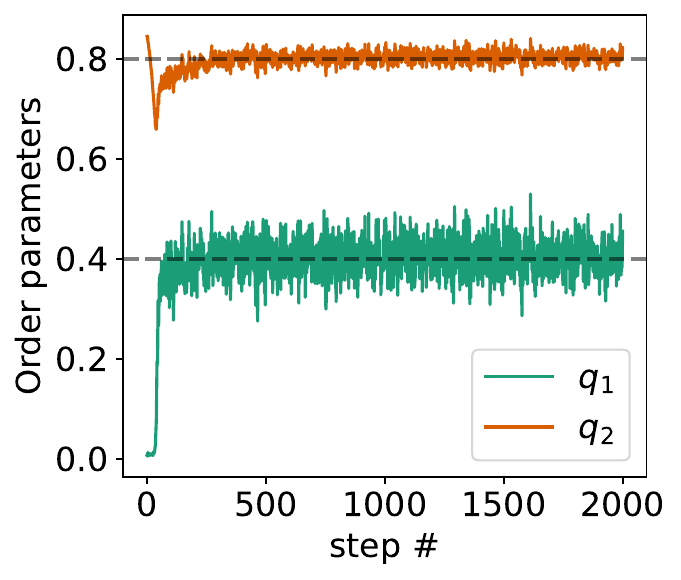}
    \includegraphics[width=0.195\linewidth]{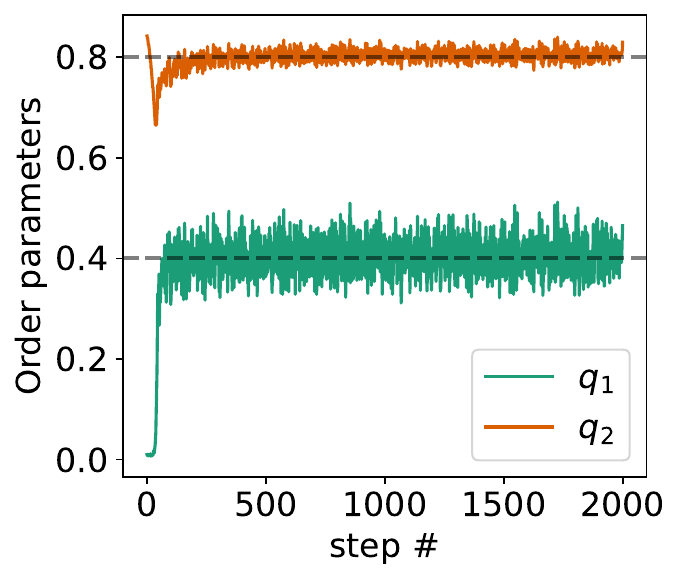}
    \includegraphics[width=0.195\linewidth]{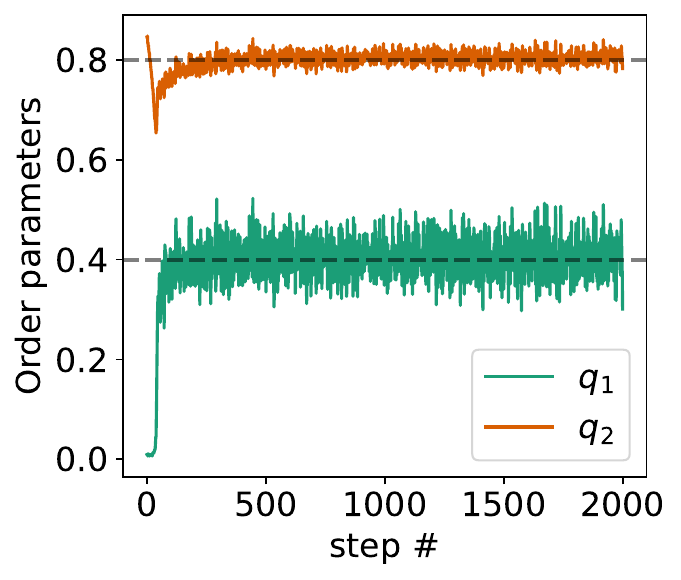}
    \includegraphics[width=0.195\linewidth]{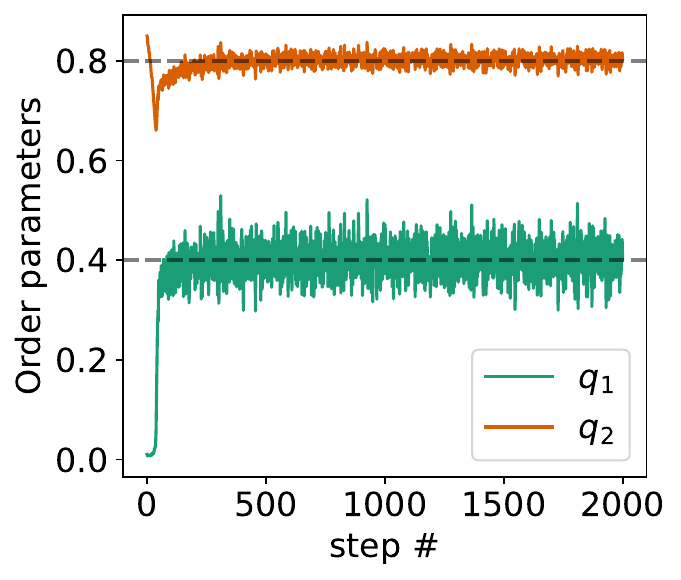}
    \includegraphics[width=0.195\linewidth]{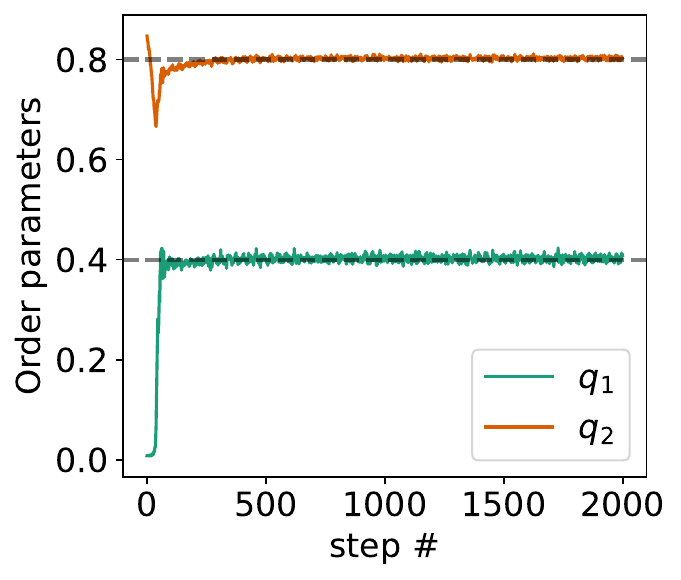}

    \includegraphics[width=0.195\linewidth]{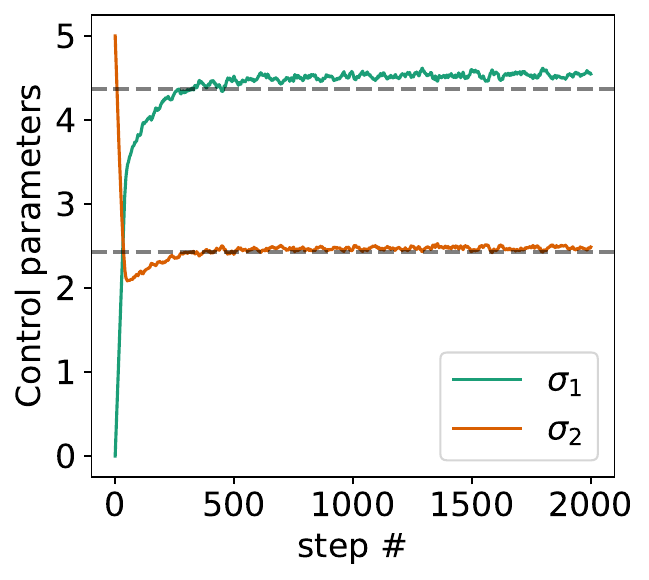}
    \includegraphics[width=0.195\linewidth]{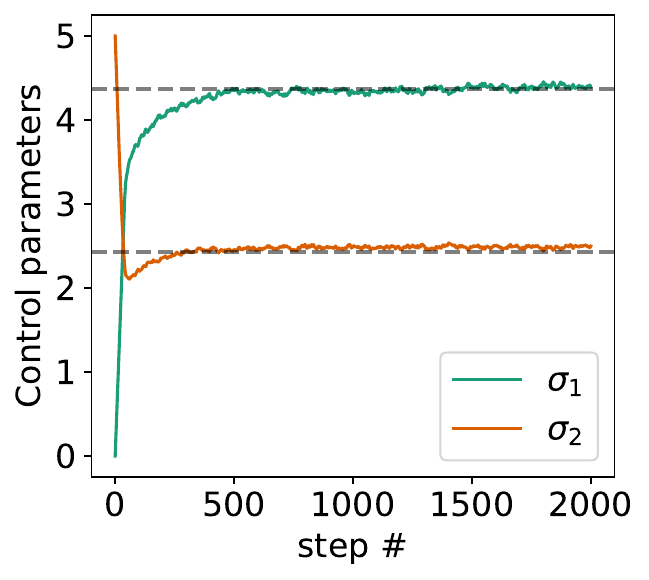}
    \includegraphics[width=0.195\linewidth]{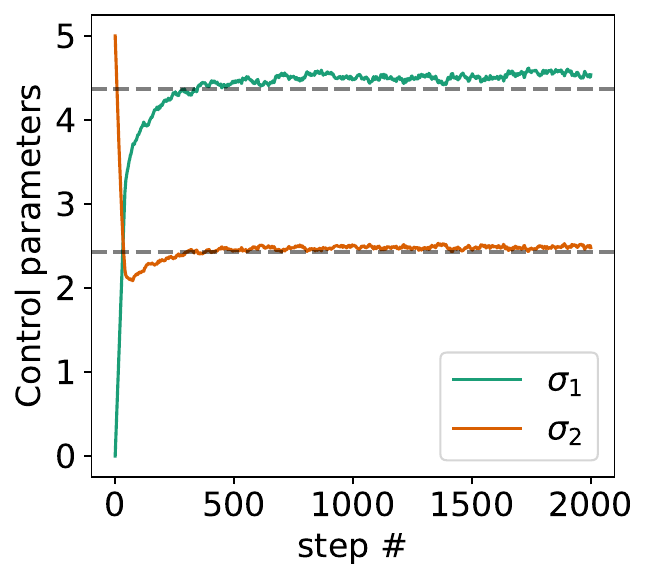}
    \includegraphics[width=0.195\linewidth]{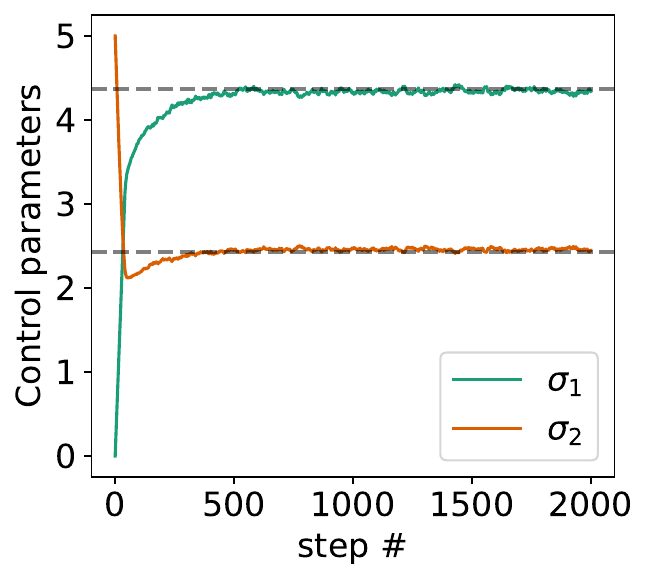}
    \includegraphics[width=0.195\linewidth]{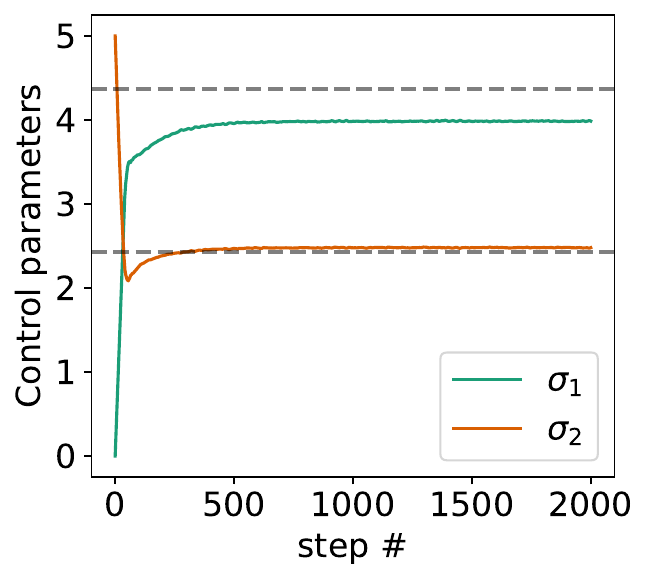}
    \caption{Evolution of order (\textit{top}) 
    and control (\textit{bottom}) parameters 
    during the adaptation process in networks with two levels ($P_1=P_2=100$).
    Five columns correspond to different seeds (i.e., independent 
    realizations of the weights and initial conditions). 
    Dashed lines denote the desired activity levels (\textit{top}) 
    and the corresponding control parameters, 
    as predicted by the mean-field equations (\textit{bottom}). 
    Other parameters: $\eta=0.2$, $\hat{q}_L=0.8$.
    }
    \label{fig:SM_adaptation_L2}
\end{figure}

\begin{figure}
    \centering
    \includegraphics[width=0.195\linewidth]{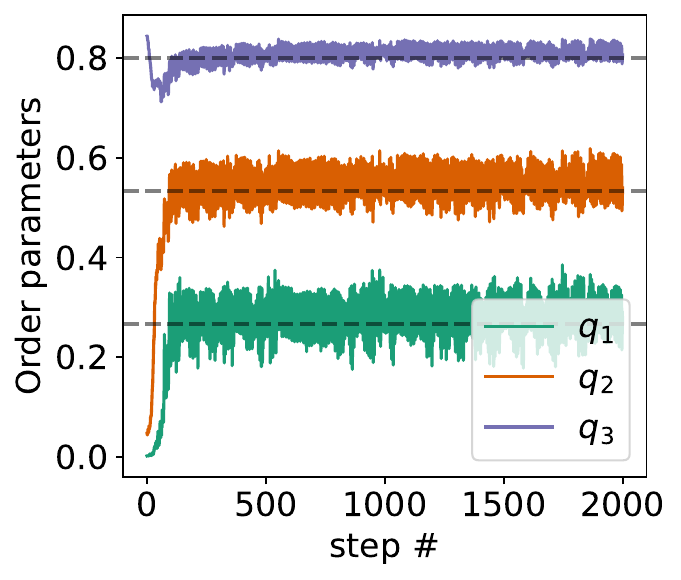}
    \includegraphics[width=0.195\linewidth]{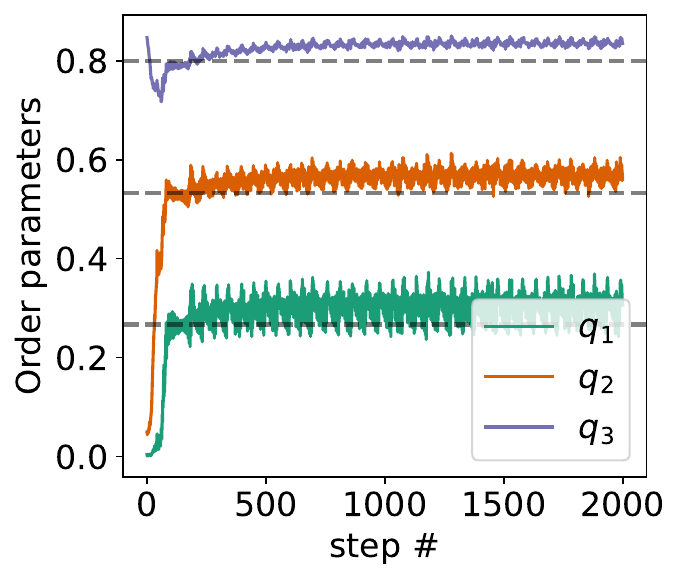}
    \includegraphics[width=0.195\linewidth]{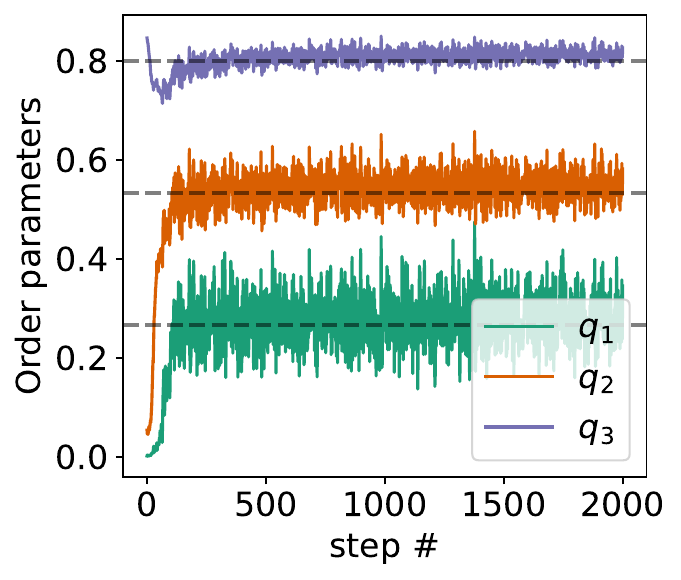}
    \includegraphics[width=0.195\linewidth]{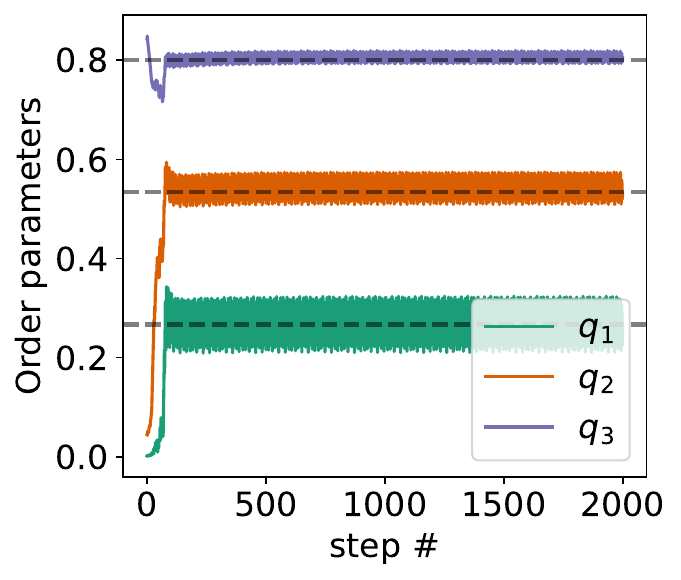}
    \includegraphics[width=0.195\linewidth]{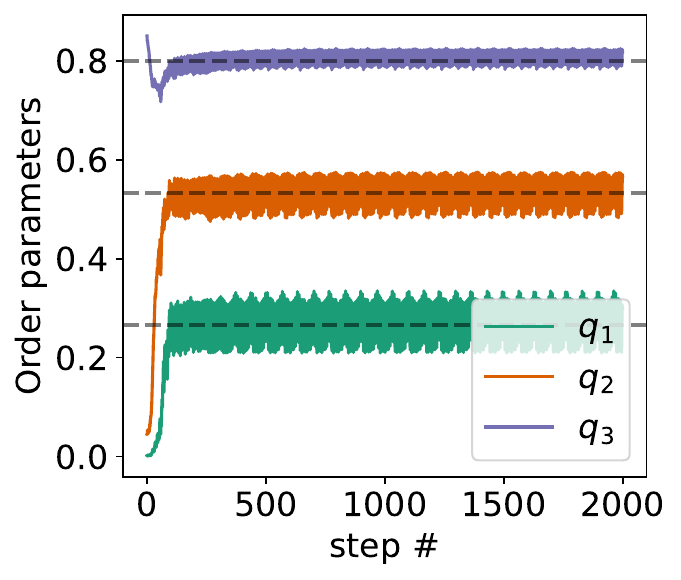}

    \includegraphics[width=0.195\linewidth]{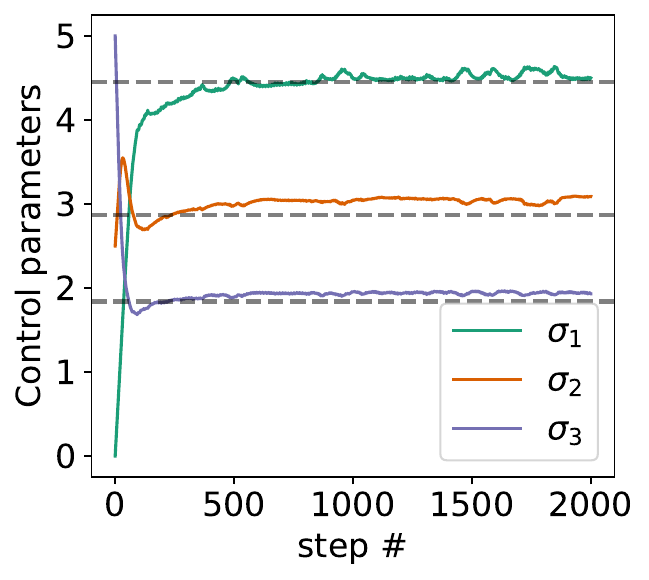}
    \includegraphics[width=0.195\linewidth]{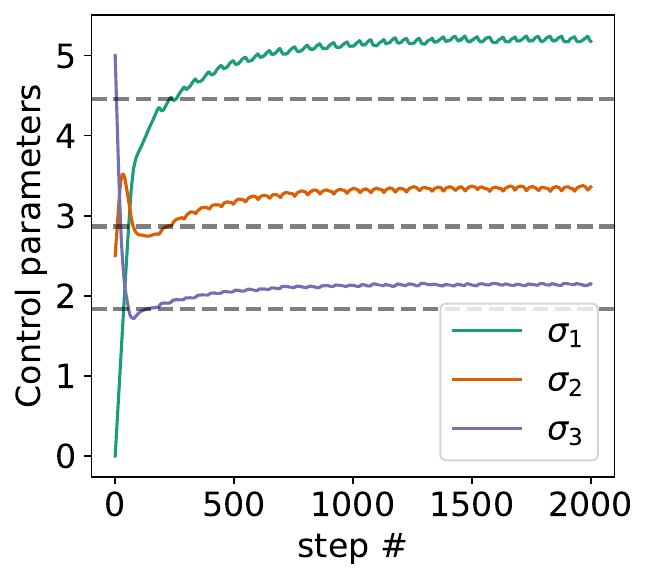}
    \includegraphics[width=0.195\linewidth]{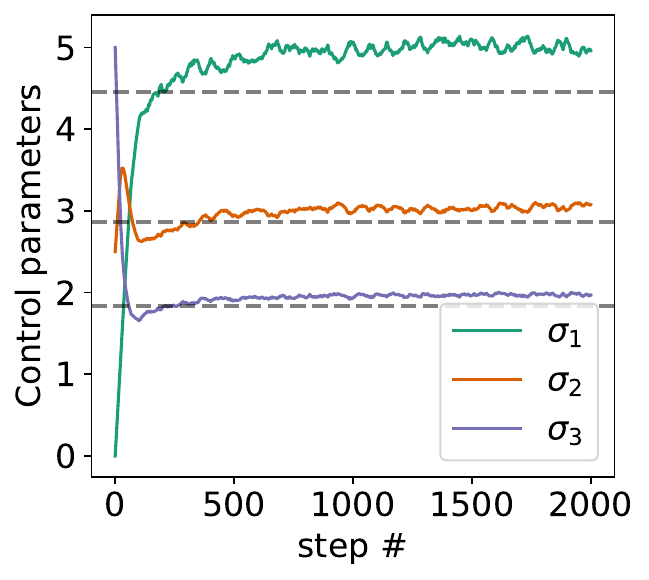}
    \includegraphics[width=0.195\linewidth]{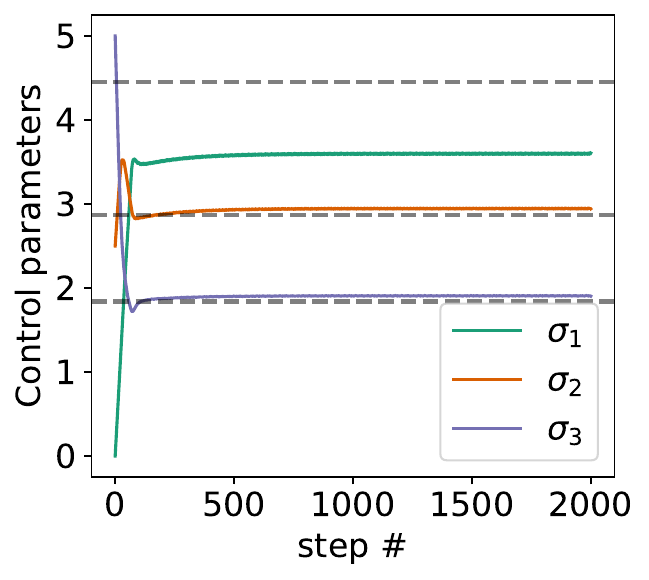}
    \includegraphics[width=0.195\linewidth]{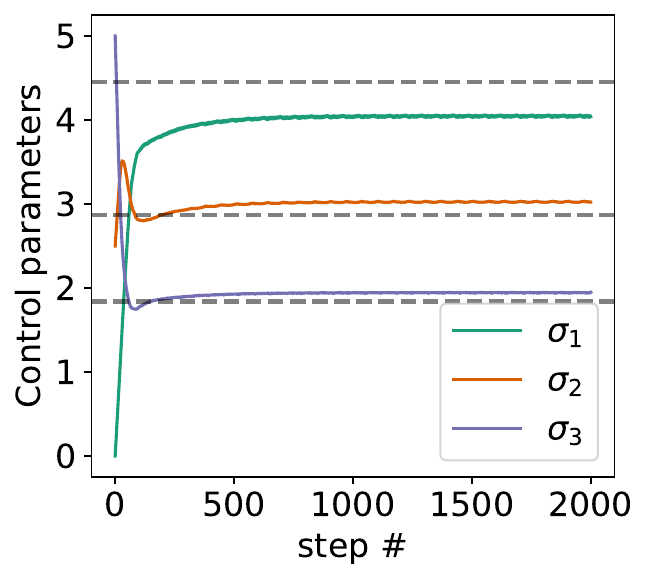}
    \caption{Same as Fig.~\ref{fig:SM_adaptation_L2} 
    but in networks with three levels 
    ($P_1=P_2=P_3=22$).
    }
    \label{fig:SM_adaptation_L3}
\end{figure}

\begin{figure}
    \centering
    \includegraphics[width=0.195\linewidth]{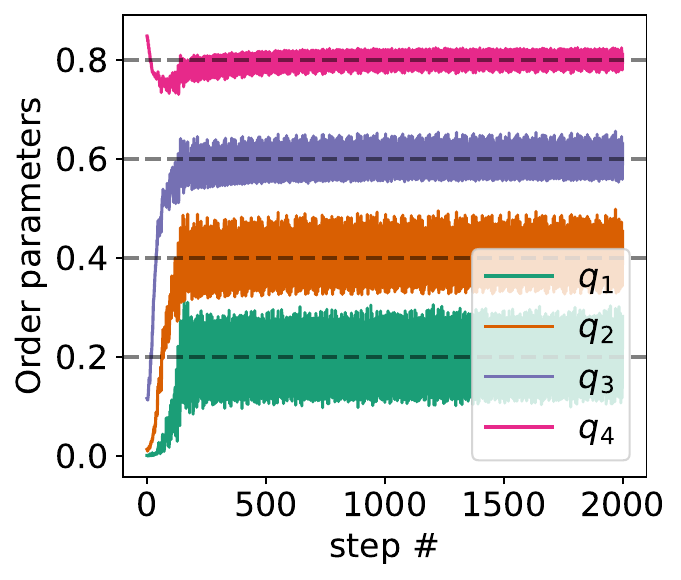}
    \includegraphics[width=0.195\linewidth]{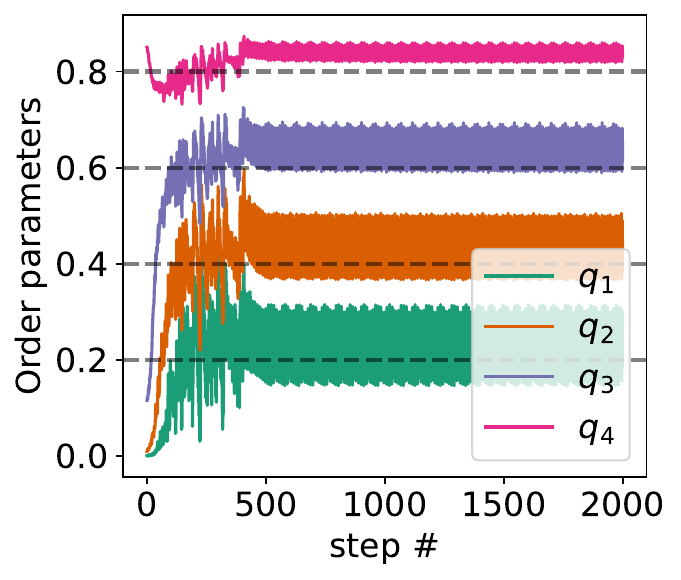}
    \includegraphics[width=0.195\linewidth]{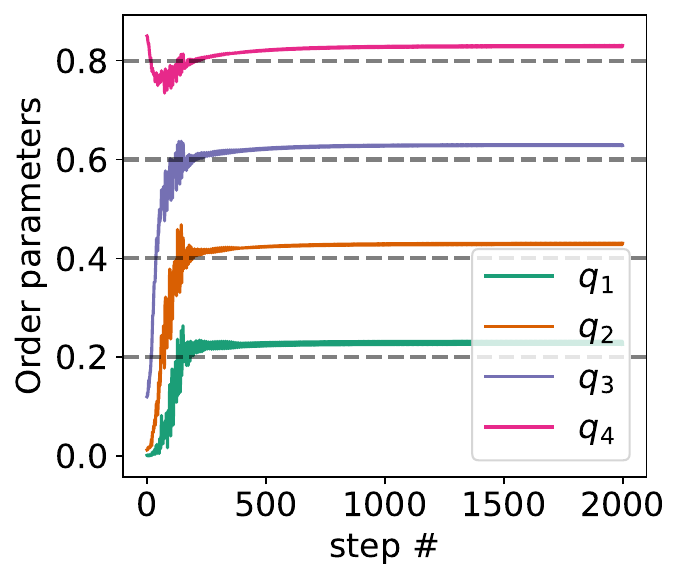}
    \includegraphics[width=0.195\linewidth]{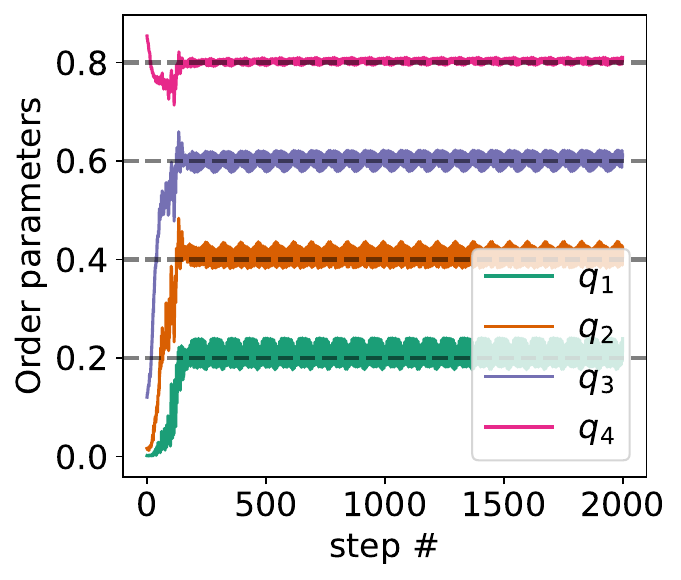}
    \includegraphics[width=0.195\linewidth]{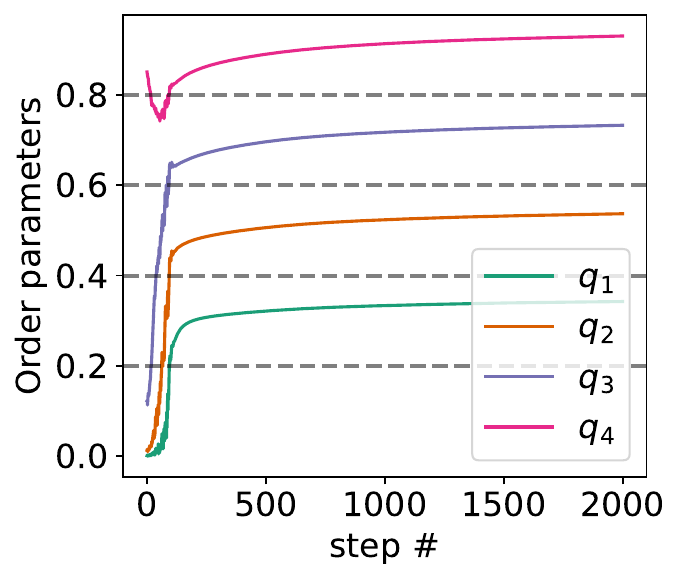}

    \includegraphics[width=0.195\linewidth]{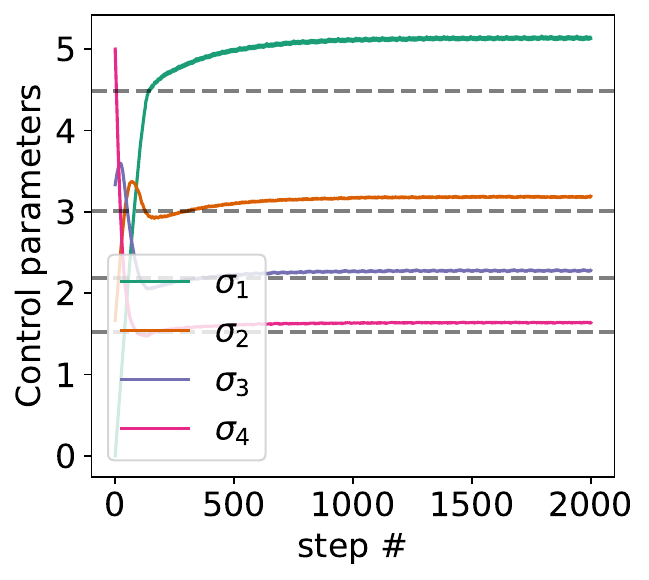}
    \includegraphics[width=0.195\linewidth]{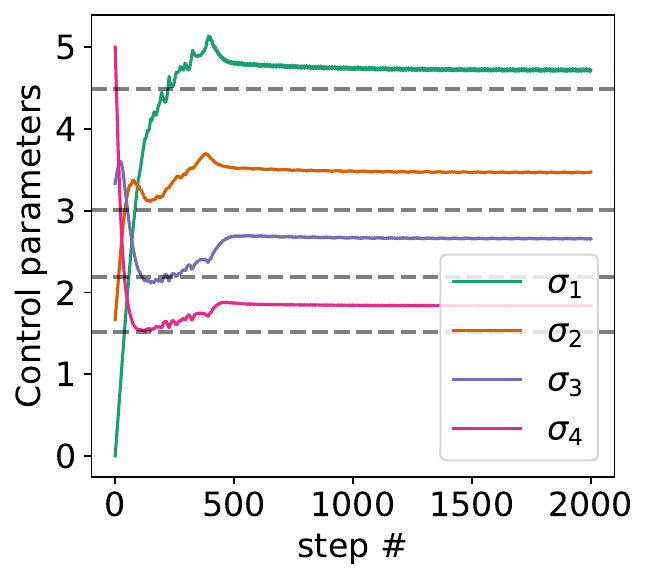}
    \includegraphics[width=0.195\linewidth]{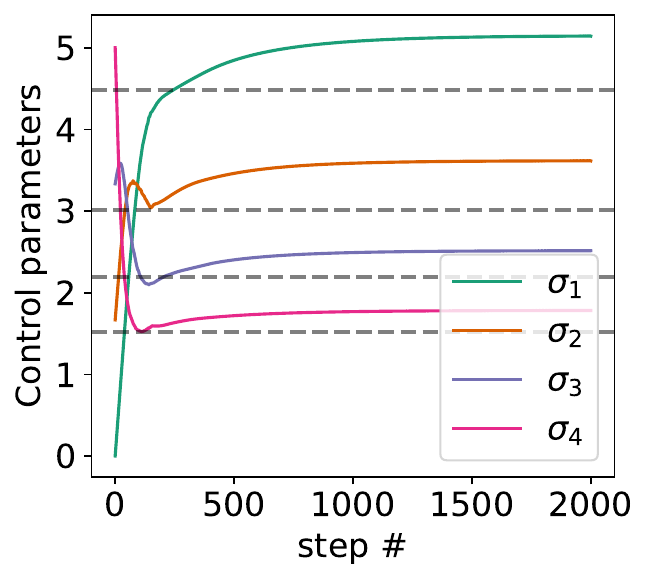}
    \includegraphics[width=0.195\linewidth]{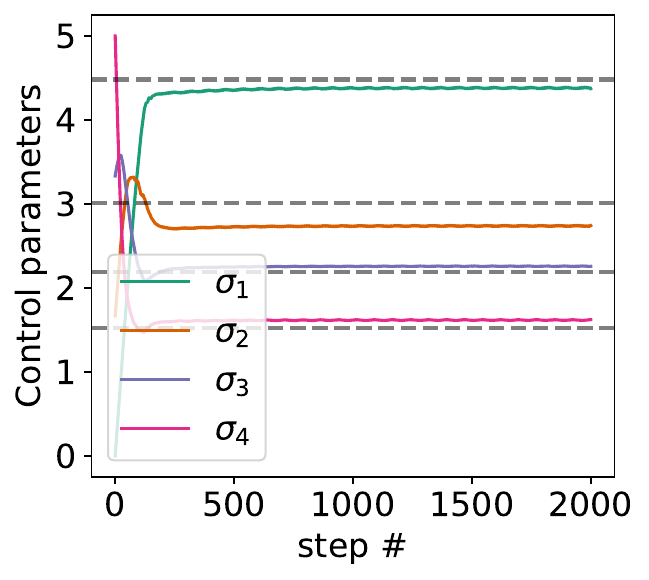}
    \includegraphics[width=0.195\linewidth]{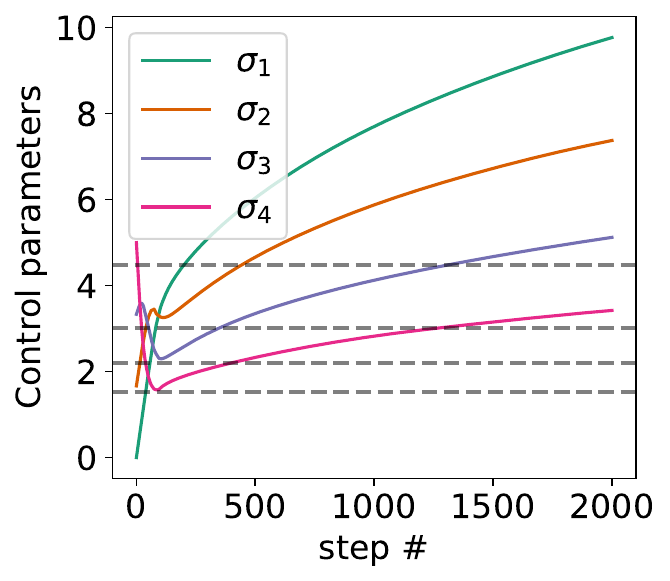}
    \caption{Same as Fig.~\ref{fig:SM_adaptation_L2} 
    but in networks with four levels ($P_1=P_2=P_3=P_4=10$).
    }
    \label{fig:SM_adaptation_L4}
\end{figure}

\begin{figure}
    \centering
    \includegraphics[width=0.195\linewidth]{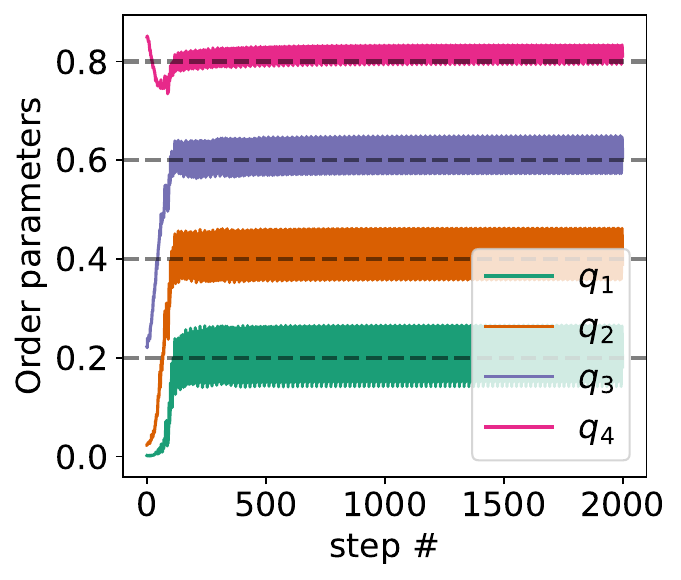}
    \includegraphics[width=0.195\linewidth]{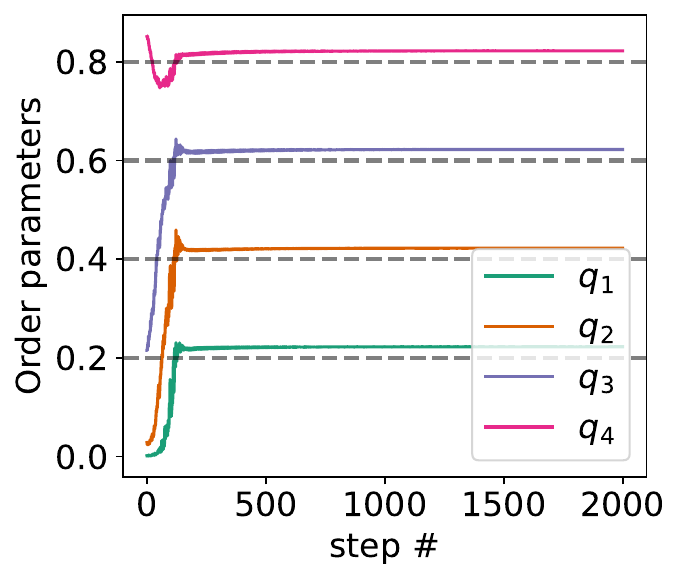}
    \includegraphics[width=0.195\linewidth]{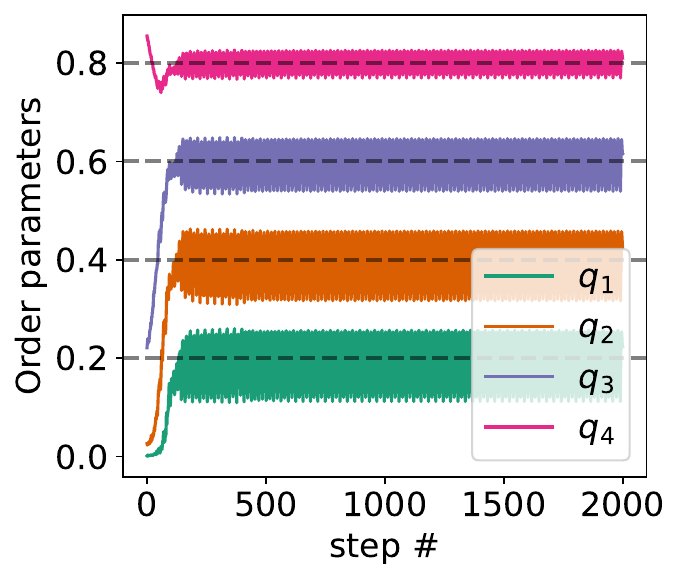}
    \includegraphics[width=0.195\linewidth]{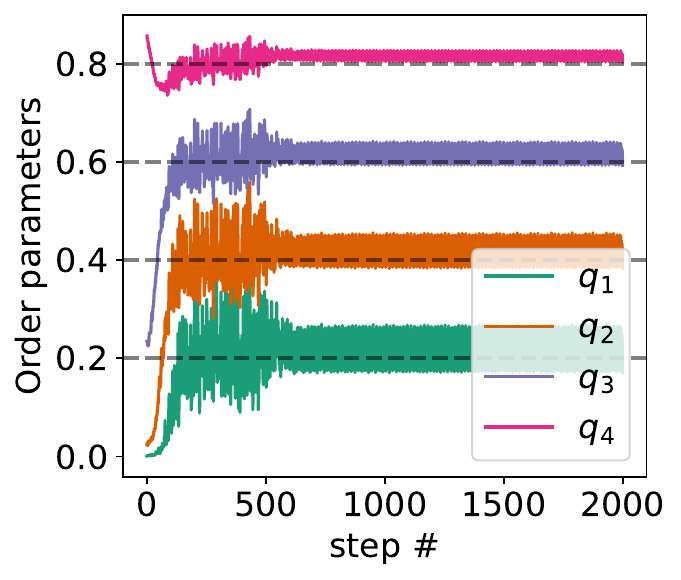}
    \includegraphics[width=0.195\linewidth]{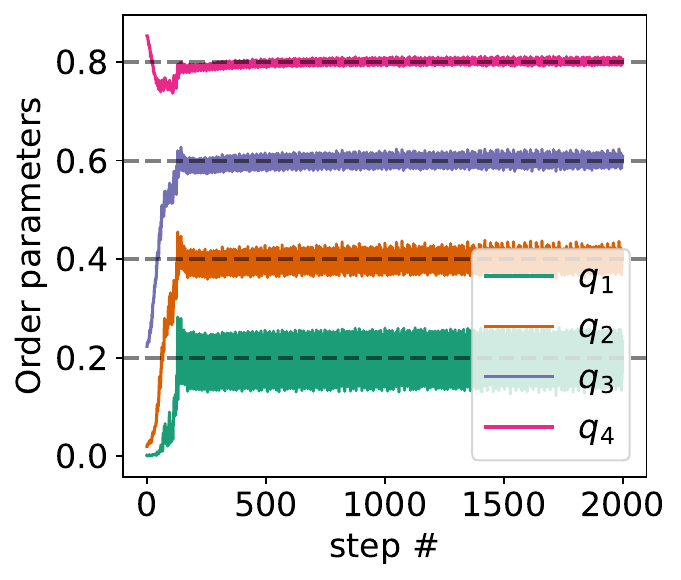}

    \includegraphics[width=0.195\linewidth]{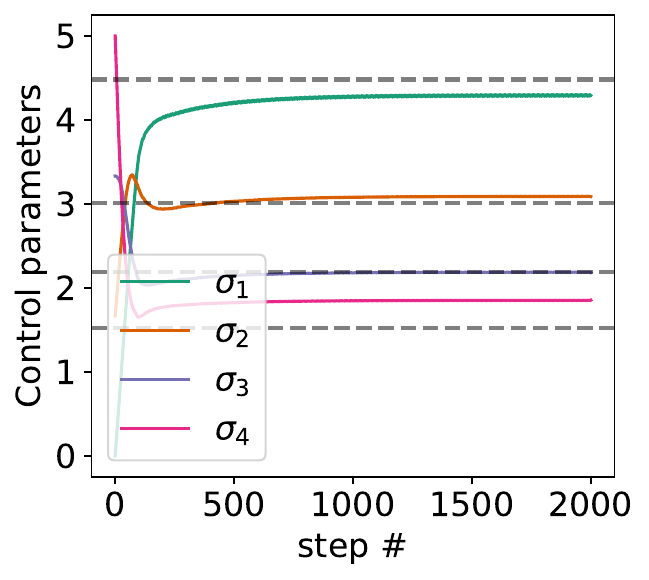}
    \includegraphics[width=0.195\linewidth]{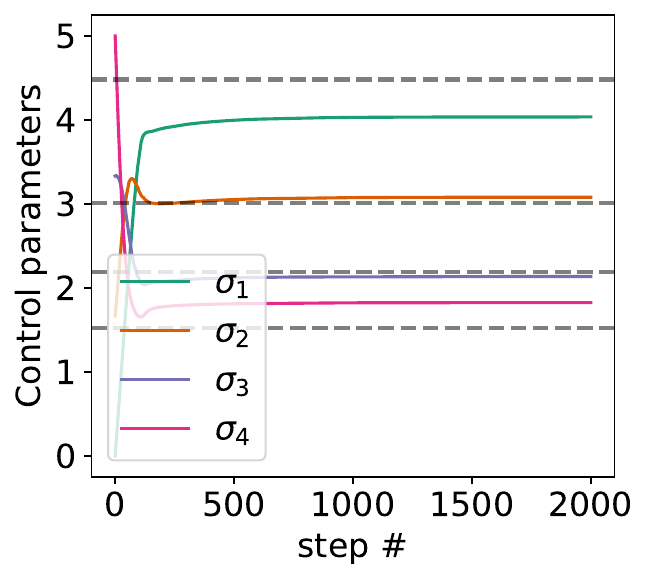}
    \includegraphics[width=0.195\linewidth]{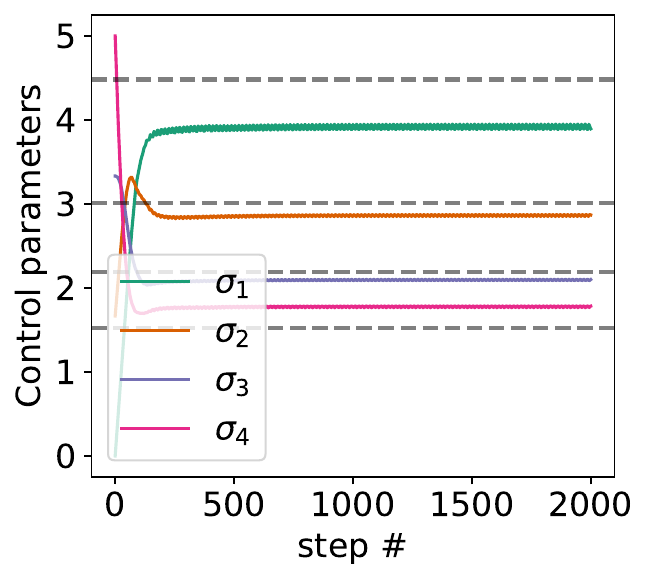}
    \includegraphics[width=0.195\linewidth]{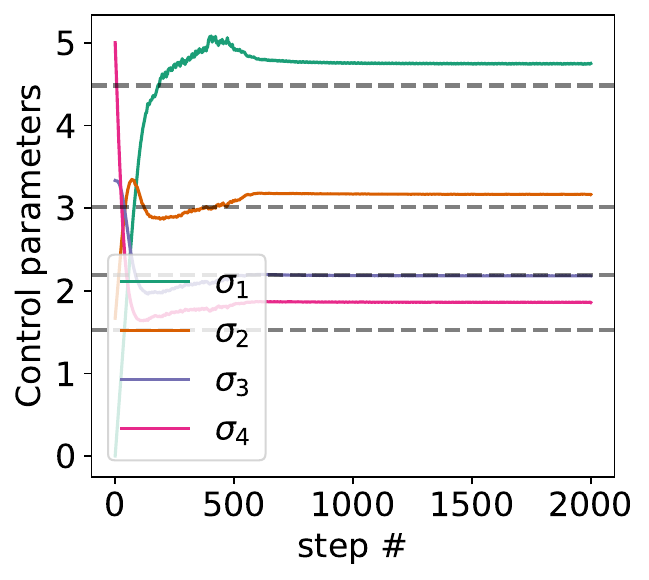}
    \includegraphics[width=0.195\linewidth]{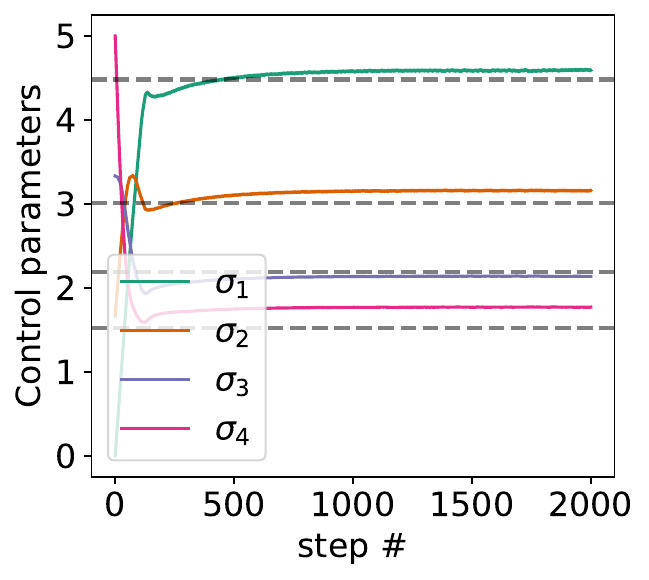}
    \caption{Same as Fig.~\ref{fig:SM_adaptation_L4} 
    but in larger networks with significantly more top-level populations 
    ($P_1=20$, $P_2=15$, $P_3=10$, $P_4=5$).
    }
    \label{fig:SM_adaptation_L4_larger}
\end{figure}

First, we test whether our adaptation algorithm brings the activity levels $q_i$ 
close to the desired values $\hat{q}_k = k q_L/L$. 
In two-level networks the order parameters quickly converge to the steady state, 
where they fluctuate around the set point (Fig.~\ref{fig:SM_adaptation_L2}, top). 
The associated values of the control parameters feature lower level of fluctuations 
whose means are close to, but may not match exactly, the values predicted 
by the mean-field theory (Fig.~\ref{fig:SM_adaptation_L2}, bottom). 
The fluctuations and the bias are expected to be finite-size effects. 
The results are similar in three-level networks (Fig.~\ref{fig:SM_adaptation_L3}). 
Occasionally, all order and control parameters seem to converge to values 
slightly larger than desired. 
This happens because in these networks $q_0$, 
which denotes the squared mean activity of the network, 
converges to a significant non-zero values. 
Our algorithm does not attempt control $q_0$, 
since in our simple formulation of the model there is no associated control parameter 
and in the thermodynamic limit $q_0$ is predicted to be equal to $0$ anyway.  
Given that $\Delta_1$ is calculated as  $q_1 - q_0$, 
the convergence of all $\Delta_i$ to $\hat{q}_L/L$ 
does not imply the convergence of all $q_i$ to $i \hat{q}_L/L$. 
Instead, $q_i$ converges to $i \hat{q}_L/L + q_0$. 
This phenomenon can lead to a failure of the algorithm when $\hat{q}_L + q_0 > 1$, 
which we occasionally observe in four-level networks when the number of populations 
is relatively small (Fig.~\ref{fig:SM_adaptation_L4}). 
However, we expect this to be a finite-size effect that does not 
persist in larger networks. 
Indeed, increasing the number of populations in four-level networks seems to 
significantly reduce the effect (Fig.~\ref{fig:SM_adaptation_L4_larger}).

\begin{figure}
    \centering
    \includegraphics[width=0.99\linewidth]{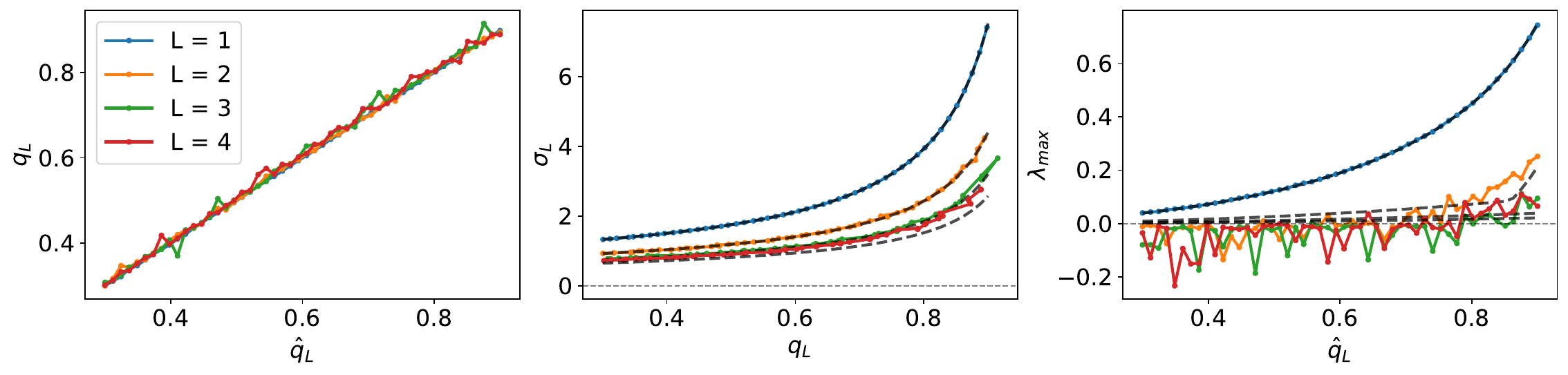}
    \caption{
    Values of $q_L$ (\textit{left}),
    $\sigma_L$ (\textit{center}),
    and the maximal Lyapunov exponent (\textit{right}) 
    at the end of the adaptation process
    as functions of $\hat{q}_L$ or $q_L$ 
    for 
    $L=1$ ($P_1=10^4$), 
    $L=2$ ($P_1=P_2=100$), 
    $L=3$ ($P_1=P_2=P_3=22$),
    and 
    $L=4$ ($P_1=20$, $P_2=P_3=10$, $P_4=5$).
    The adaptation process runs for $1000$ steps, 
    after which control parameters are frozen and the simulation is restarted 
    from a new random initial condition for $700$ steps. 
    The last $350$ steps are used for estimating the order parameters $q_i$, 
    whereas the last $200$ steps are used for estimating $\lambda_{max}$. 
    }
    \label{fig:SM_adaptation_summary_different_qL}
\end{figure}

We repeat this procedure for various values of $\hat{q}_L$, 
as summarized in Fig.~\ref{fig:SM_adaptation_summary_different_qL}. 
As expected, the adaptation process drives $q_L$ close to $\hat{q}_L$ 
and, in most cases, 
brings $\sigma_L$ to values predicted by the mean-field theory. 
The only exception is for $L=4$, where the $\sigma_L$ resulting from the adaptation remains significantly larger than the mean-field prediction, presumably due to finite-size effects.
Given that our algorithm successfully balances activity across levels, 
we expect it to position the system near the edge of chaos. 
The right panel of Fig.~\ref{fig:SM_adaptation_summary_different_qL} 
confirms this, showing that the maximal Lyapunov exponents remain 
significantly closer to $0$ for modular networks ($L=2$ and $L=3$) 
compared to homogeneous networks ($L=1$). 
The estimated values of $\lambda_{max}$ align with the mean-field predictions, 
although modular networks exhibit relatively large 
variability across realizations.
Despite these finite-size effects, we conclude that balancing activity 
across levels in modular networks places neural dynamics in the 
vicinity of the edge of chaos. 

\end{document}